\documentclass[final,twocolumn,12p]{elsarticle}

\usepackage{balance}
\usepackage{multirow}
\usepackage{blindtext, graphicx}
\usepackage{boldline}
\usepackage{subfig} 
\usepackage{ragged2e}
\usepackage[para,online,flushleft]{threeparttable}

\usepackage[tikz]{bclogo}
\newenvironment{RQ}[1]%
{\noindent\begin{minipage}[c]{\linewidth}%
\begin{bclogo}[couleur=gray!20,%
                arrondi=0.1,%
                logo=\bclampe,%
                ombre=true]{~#1}}%
{\end{bclogo}\end{minipage}}

\usepackage{enumitem}
\setlist[itemize]{leftmargin=*}
\setlist[enumerate]{leftmargin=*}
\usepackage{makecell}
\usepackage[linesnumbered,ruled,vlined]{algorithm2e}
\setlist{nolistsep} 

\SetKwProg{Fn}{Function}{}{}

\setlist[1]{itemsep=0pt}

\usepackage{amsmath}

\usepackage{times}
\usepackage{colortbl}
\usepackage{tikz}

\makeatletter
\let\th@plain\relax
\makeatother

\definecolor{Gray}{rgb}{0.88,1,1}
\definecolor{Gray}{gray}{0.85}
\definecolor{lightgray}{gray}{0.8}
\usepackage[framed]{ntheorem}
\usepackage{framed}
\usepackage{tikz}
\usetikzlibrary{shadows}
\theoremclass{Lesson}
\theoremstyle{break}

\tikzstyle{thmbox} = [rectangle, rounded corners, draw=black,
fill=Gray!20,  drop shadow={fill=black, opacity=1}]

\newshadedtheorem{lesson}{Answer to RQ}

\usepackage{natbib}
\usepackage[colorlinks=true]{hyperref} 

\journal{Expert Systems with Applications}

\begin{document}
\begin{frontmatter}

\title{FAST$^2$: an Intelligent Assistant for Finding Relevant Papers}

\author{Zhe Yu}
\ead{zyu9@ncsu.edu}
\author{Tim Menzies}
\ead{timm@ieee.org}

\address{Department of Computer Science, North Carolina State University, Raleigh, NC, USA}

\begin{abstract}
Literature reviews are essential for any researcher trying to keep up to date with the burgeoning software engineering literature. Finding relevant papers can be hard due to the huge amount of candidates provided by search. FAST$^2$ is a novel tool for assisting the researchers to find the next promising paper to read. This paper describes FAST$^2$ and tests it on four large systematic literature review datasets.
We show that FAST$^2$ robustly optimizes the human effort to find most (95\%) of the relevant software engineering papers while also compensating for the errors made by humans during the review process.
The effectiveness of FAST$^2$ can be attributed to three key innovations:
(1) a novel way of applying external domain knowledge (a simple two or three keyword search) to  guide the initial selection of papers---which helps to find relevant research papers faster with less variances;
(2) an estimator of the number of remaining relevant papers yet to be found---which helps the reviewer decide when to stop the review;
(3) a novel human error correction algorithm---which corrects a majority of human misclassifications (labeling relevant papers as non-relevant or vice versa) without imposing too much extra human effort.
\end{abstract}


\end{frontmatter}

\section{Introduction}
\label{sect: Introduction}

\begin{table*}[!htb]
\caption{Statistics from the literature review case studies in this paper}
\label{tab: stats}
\begin{center}
\begin{threeparttable}
\small

\setlength\tabcolsep{7pt}
\begin{tabular}{ l|c|c|c|c|c }
   Dataset & Topic & \makecell{Original Title} & \makecell{\#candidate   \\ papers $|E|$}& \makecell{\#relevant  \\ papers $|R|$}& \makecell{Prevalence \\$(|R|/|E|)\%$} \\
  \hline
  Wahono &  Defect prediction & \makecell{A systematic literature review \\of software  defect prediction: \\research trends,  datasets, \\ methods and frameworks} & 7002 & 62 & 0.9\% \\
  \hline
  Hall & Defect prediction & \makecell{A systematic review of theory \\use in  studies investigating the\\  motivations of software engineers} & 8911 & 104 & 1.2\%\\
  \hline
  Radjenovi{\'c} &\makecell{ Defect prediction\\ metrics} & \makecell{Software fault prediction metrics:\\ A systematic literature review} & 6000 & 48 & 0.8\%\\
  \hline
  Kitchenham & Literature review & \makecell{A systematic review of systematic\\ review  process research in SE} & 1704 & 45 & 2.6\% \\
  \hline
\end{tabular}
\begin{tablenotes}\small
Datasets used in the authors' prior work~\citep{yu2018finding}. The first three datasets are generated by reverse engineering from the original publications (using the same search string to collect candidate papers and treating the final inclusion list as ground truth for relevant papers). 
\end{tablenotes}
\end{threeparttable}

\end{center}
\end{table*}

This article presents and assesses an automatic assistant for researchers seeking   research papers  relevant to their particular  topic. The goal of this assistant, called FAST$^2$, is to reduce the effort required for such a search thereby  (a)~enabling researchers to find more relevant papers faster;
thus (b)~allowing  the contribution of more researchers to be wildly recognized. Many such assistants have been previously proposed~\citep{cormack2014evaluation,wallace2010semi,miwa2014reducing,yu2018finding}, but those prior works did not fully address three core problems solved by FAST$^2$ (see our research questions, listed on the next page). 

Why is it important to reduce the effort associated with
literature reviews?
Given the prevalence of tools like SCOPUS, Google Scholar, ACM Portal, IEEE Xplorer, Science Direct, etc., it is a relatively simple task to find a few relevant papers for any particular
research query. 
However, what if the goal is not to find a {\em few} papers, but instead to find {\em most} of the relevant papers?  Such broad searches are very commonly conducted by:
\begin{itemize}
\item
Researchers exploring a new area;
\item
Researchers writing papers for peer review, to ensure reviewers will not reject a paper since it omits important related work;
\item
Researchers conducting a systematic literature review on a specific field to learn about and summarize the latest developments. 
\end{itemize}
Literature reviews can
be extremely labor intensive due to the low prevalence of relevant papers. Here in this paper, we take the example of systematic literature reviews to demonstrate this problem. 
For example,
{\em systematic literature reviews} are the primary method for aggregating evidence in evidence-based software engineering~\citep{keele2007guidelines}. In such reviews, researchers thoroughly analyze all the research papers they can find to synthesize answers to some specific research questions. One specific step in systematic literature review, which is called primary study selection, is to find most (if not all) of the relevant papers to the research questions. This step is identified as one of the most difficult and time consuming steps in systematic literature review~\citep{carver2013identifying} as it requires humans to read and classify thousands of candidate papers from a search result. Table~\ref{tab: stats} shows 
  four systematic literature reviews where thousands
of papers were reviewed before revealing just a few dozen relevant papers. 
\citet{shemilt2016use} estimated that assessing one
paper for relevancy takes at least one minute. Assuming
25 hours per week for this onerous task,    the studies of Table~\ref{tab: stats}  would take  16 weeks to complete (in total). Therefore, reducing the human efforts required in this primary study selection step is thus critical for enabling researchers conducting systematic literature reviews more frequently.


Many researchers have tried reducing the effort associated with literature reviews~\citep{malheiros2007visual,bowes2012slurp,jalali2012systematic,wohlin2014guidelines,o2015using,paynter2016epc,cohen2006reducing,adeva2014automatic,liu2016comparative,ros2017machine}. Prior work from the authors~\citep{yu2018finding} focused on retrieving most relevant papers with least effort and found that active learning is the best way to achieve that. In the state of the art active learning approaches~\citep{cormack2014evaluation,wallace2010semi,miwa2014reducing}, a support vector machine (SVM) is updated whenever a human reviewer decided if a paper is relevant/non-relevant to a specific research question. The active learner then reflect over the updated SVM to select which paper is most informative to read next. In this active learning manner, human review efforts are focused on the papers that either is most likely to be relevant or can improve the current classifier most. Furthermore, advantages from different active learning approaches are adopted to derive a new approach FASTREAD~\citep{yu2018finding}, which outperformed the previous state of the art approaches~\citep{cormack2014evaluation,wallace2010semi,miwa2014reducing}.
 
While a useful tool, FASTREAD does not address three important research questions (that are resolved in this paper):

{\bf RQ1: ``How to start?''}; i.e., how to control initial paper selection. 
The resulting incremental learners
can vary wildly, depending on the initial selection of examples. This is important since,
 as shown below in Section~\ref{sect: limit}, 
 a poor initial selection of papers
can greatly increase the number of papers that must be read.   But as shown in this paper: 

\begin{RQ}{How to start?}
The first relevant paper can be identified earlier, robustly, by applying a little domain knowledge to guide the initial sampling. 
\end{RQ}

{\bf RQ2:  ``When to stop?''}; i.e., how to know when can the review be safely stopped. 
While there is probably always one more relevant paper to find, it would be useful to know when
most papers have been found 
 (say, 95\%  of them~\citep{cohen2011performance}). Without the knowledge of how many relevant papers that are left to be found, researchers might either:
 \begin{itemize}
\item
Stop too early, thus missing many relevant papers.
\item
Stop too late, causing unnecessary further reading, even after all relevant items are find. 
\end{itemize}
This paper shows that the rate at which incremental SVM tools finds papers follow a simple mathematical relationship. Hence:

 \begin{RQ}{``When to Stop?''}
During the review process, the current achieved recall of relevant papers can be accurately estimated through a semi-supervised logistic regressor. Therefore it is possible to determine when a target recall (say, 95\%), has been reached.
\end{RQ}
 We  show,  in Section~\ref{sect: When to Stop: Results} that
 our stopping rule is more accurate than other state of the art stopping criteria~\citep{wallace2013active}.

{\bf RQ3:  ``How to correct?''}. Human reviewers
are not perfect, and sometimes they will label relevant
papers as non-relevant, and vice versa. Practical
tools for support literature reviews must be able to recognize and repair incorrect labeling. We show here that:

\begin{RQ}{How to correct?}
The human labeling errors can be efficiently identified and corrected by periodically rechecking few of the labeled papers, whose labels the active learner disagrees most on.
\end{RQ}

This paper is structured as follows.
The rest of this section states our contributions; connections to prior work; and some caveats. After that,  background notes are presented selecting relevant papers in Section~\ref{sect: Background}, techniques and experimental results for the \emph{how to start} problem in Section~\ref{sect: How to Start}, answering to the \emph{when to stop} problem in Section~\ref{sect: When to Stop}, and results for \emph{How to correct} problem when human errors are taken into consideration in Section~\ref{sect: How to correct}. Section~\ref{sect: discussion} describes the overall tool FAST$^2$ as a combination of all the techniques from Section~\ref{sect: How to Start} to Section~\ref{sect: How to correct} and discusses the threats to validity. Conclusions are provided in Section~\ref{sect: Conclusion}.

Note that all our case studies are systematic literature reviews from the field of software engineering. We use that data since that is our ``home''
domain and it seems more responsible to draw data from an area that we know most about. That said, there is nothing in principle stopping the reader
from applying these methods to their own home domains. To that end, we offer a full reproduction package with all this study's code
and data~\citep{fastread}.

\subsection{Contributions}
This paper offers:
\begin{enumerate}
\item
A starting tactic for literature reviews to find the first relevant paper faster, robustly.
\item
An early stopping rule which, for our test data,
is far more accurate than other state of the art stopping criteria.
\item
An error correcting strategy that mitigates human errors in literature reviews.
\item
A software  tool called FAST$^2$ that implements the above
three points, thus addressing three open issues in prior fast reading tools, i.e. \emph{how to start}, \emph{when to stop},  \emph{how to correct}.
\item
A reproduction package with all this study's code and data~\citep{fastread}.
\end{enumerate}

\begin{figure*}[!t]
\begin{center}
\includegraphics[width=\textwidth]{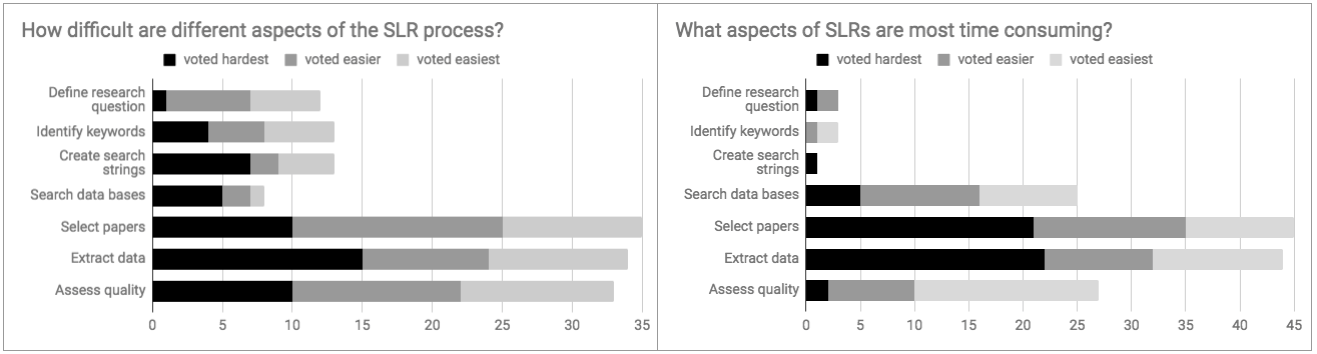}
\end{center}
\caption{This paper explores tools for reducing the effort and difficulty associated with the {\em select papers} 
phase of systematic literature reviews.
We do this since \citet{carver2013identifying} polled researchers who have conducted literature reviews.
According to their votes, {\em select papers} was one of the hardest, most time-consuming tasks
in the literature review process. For tools that support other aspects of SLRs such as searching databases and extracting data, see summaries from \citet{marshall2015tools}. }\label{fig:barrier}
\end{figure*}

\subsection{Connection to Prior Work}

This paper is  a significant improvement over FASTREAD, the
tool proposed in our prior work~\citep{yu2018finding} on
automatic support for literature reviews. In this prior work~\citep{yu2018finding}, the authors 
\begin{enumerate}
\item
Analyzed the shortcomings of existing systematic review tools, e.g. Abstrakr~\footnote{http://abstrackr.cebm.brown.edu}~\citep{wallace2012deploying}, EPPI-Reviewer~\footnote{http://eppi.ioe.ac.uk/cms/er4/}~\citep{thomas2010eppi}, Rayaan~\footnote{http://rayyan.qcri.org/}~\citep{Ouzzani2016}. 
\item
Compared different approaches, such as search-query based methods~\citep{zhang2011empirical,Umemoto2016ScentBar}, reference-based methods~\citep{jalali2012systematic,felizardo2016using,wohlin2014guidelines}, supervised learning~\citep{cohen2006reducing,adeva2014automatic}, semi-supervised learning~\citep{liu2016comparative}, unsupervised learning~\citep{malheiros2007visual}, and found that active learning is the most efficient in reducing the cost of primary study selection.
\item
Constructed three software engineering systematic literature review datasets by reverse engineering from the original publications. 
\item
Mixed and matched the state of the art active learning approaches~\citep{wallace2010semi,miwa2014reducing,cormack2014evaluation} to generate 
32 possible combinations. 
\item
The resulting best method FASTREAD had top rank performance across all datasets and significantly outperformed the three existing state of the art methods.
\end{enumerate}
More details on FASTREAD will be provided in Section~\ref{sect: FASTREAD}. While FASTREAD was the state of the art approach for selecting which papers to
read next, it offered no support for the {\bf RQ1, RQ2, RQ3} issues that are solved in this paper.

\subsection{Caveats}

One caveat for the following results is that all our test data comes from the software engineering literature.
We constrained ourselves to that domain since we have 
personnel contacts with many researchers performing  literature reviews in that domain. This constraint had many advantages, including our ability to access more test data. 
Note that  there is nothing in principle from applying our methods
to other domains. 

Another important caveat is that we say that we are ``reducing the effort of literature reviews'', we mean ``reduce the effort of the initial skim of the papers''.  Formally, in the terminology of Kitchenham's systematic literature review (SLR)~\citep{kitchenham2004procedures} framework, this ``initial skim'' is the {\em primary study selection} phase of a literature review.
We focus on this phase of the literature review process for two reasons:
\begin{itemize}
\item
Prior work by Carver and others~\citep{carver2013identifying,hassler2014outcomes,hassler2016identification} showed that primary study selection
 is one of the slowest parts of the entire literature review process, as shown in Figure~\ref{fig:barrier}. 
\item
There are well-established tools to handle much of the rest of the SLR process. From example, \citet{marshall2015tools} describe tools to assist in  {\em searching} for papers then  {\em extracting} papers from multiple databases and  {\em assessing} their ``quality'' (i.e. their relevance to the query). But a large missing piece in the current toolchain for SLRs are supports for {\em selecting} a small number of relevant papers from within the larger set of papers found by {\em searching} the databases. 
 \end{itemize}
Another important caveat is that 
we are careful to say that our methods can find {\em many} relevant papers, rather than {\em all} relevant papers.
Our goal is to offer engineering principles for literature reviews, and engineering is the discipline that delivers acceptable products in reasonable time. Hence, like many other researchers in this area, our success criteria is
``work saved oversampling at 95\% recall'' (WSS@95)~\citep{cohen2011performance}.
If researchers hope to find {\em all} relevant papers, then they should 
(a)~ignore the methods of this paper and (b)~allocate a very long time to their literature reviews.

\section{Background}
\label{sect: Background}

\subsection{Literature Reviews in Different Domains}
Selecting which technical papers to read is a task that is relevant and useful for many domains. For example:
\begin{itemize}
\item
In legal reasoning, attorneys are paid to review millions of documents trying to find evidence to some case. Tools to support this process
are  referred to as {\em electronic discovery}~\citep{grossman2013,cormack2014evaluation,cormack2015autonomy,cormack2016scalability}.
\item
In evidence-based medicine, researchers review medical publications to gather evidence for support of a certain medical practice or phenomenon. The selection of related medical publications among thousands of candidates returned by some search string is called {\em citation screening}. Text mining methods are also applied to reduce the review effort in citation screening~\citep{miwa2014reducing,wallace2010semi,wallace2010active,wallace2011should,wallace2013active,wallace2012deploying,wallace2013modernizing}.
\item
In software engineering, Kitchenham~\emph{et al.} recommend 
{\em systematic literature reviews} (SLRs) to be standard procedure in research~\citep{kitchenham2004evidence,keele2007guidelines}. In systematic literature reviews, the process of selecting relevant papers is referred to as primary study selection when SE researchers review titles, abstracts, sometimes full texts of candidate research papers to find the ones that are relevant to their research questions~\citep{kitchenham2004procedures,keele2007guidelines}.
\end{itemize}
For the most part,   SLRs in SE are a mostly
manual process
 (evidence: none of the   researchers surveyed in Fig~\ref{fig:barrier} 
 used any automatic tools to reduce the search space for their reading).
 Manual SLRs can be too slow to complete or repeat. Hence, the
 rest of this paper explores methods to reduce the effort associated with literature reviews.

\subsection{Automatic Tool Support}

Formally, the problem explored in this paper can be expressed using the nomenclature of Table~\ref{tab: problem}:
\begin{itemize}
\item
Starting with $L=\emptyset$ papers;
\item
Prioritize which papers to be reviewed so as to...
\item
Maximize $|L_R|$ (the number of relevant papers discovered)...
\item
While minimizing $|L|$ (the number of papers reviewed).
\end{itemize}
Tools that address this problem
can be divided as follows:

\emph{Search-query based} that involves query expansion/rewriting   based on user feedback~\citep{Umemoto2016ScentBar}. This is the process 
performed by any user  
as they run one query and reflect on the relevancy/irrelevancy of the papers returned by (e.g.) a Google Scholar query.
In this search-query based approach, users struggle
to rewrite their query in order find fewer non-relevant papers.
While a simple method to manually rewrite search queries, its performance in selecting relevant papers is outperformed by the abstract-based methods described below~\citep{clef2017,task2}.

\begin{table}[!t]
\caption{Problem description}
\label{tab: problem}
\begin{tabular}{l@{~}p{2.3in}}
\rowcolor{gray!10} 
$E$: & the set of all candidate papers\\\rowcolor{gray!10}&(returned from search).\\\rowcolor{gray!10} 
$R\subset E$: &  set of ground truth relevant papers. \\\rowcolor{gray!10} 
$I=E\setminus R$: &  set of ground truth non-relevant papers\\\rowcolor{gray!10} 
$L\subset E$: &  set of labeled/reviewed papers,\\\rowcolor{gray!10}& each review reveals whether a paper\\\rowcolor{gray!10} &  is included or not.\\\rowcolor{gray!10} 
$\neg L=E\setminus L$: &  set of unlabeled/unreviewed papers.\\\rowcolor{gray!10} 
$L_R=L\cap R$: &   identified relevant (included) papers.\\\rowcolor{gray!10} 
$L_I=L\cap I$: &   identified non-relevant (excluded)  papers.
\end{tabular}
\end{table}

\emph{Reference based} methods such as   ``snowballing'' exploit citation links between research papers. Starting with papers known to be relevant, ``forward'' and ``backward'' snowballing chases new relevant papers through papers cite the known relevant ones and references of known relevant ones, respectively~\citep{jalali2012systematic,felizardo2016using,wohlin2014guidelines}. While a straightforward method,  reference-based methods suffer from two  disadvantages---a) the cost for extracting references and forming a citation link graph ( full papers have to be
investigated before they go into the snowballing procedure~\citep{wohlin2014guidelines}); b) precision and recall vary a lot across different studies (Wohlin reported in one of his studies as precision to be 6.8\%~\citep{wohlin2014guidelines} while in another study as precision to be $15/64=23.4\%$~\citep{jalali2012systematic}).

\emph{Abstract based} methods use text from
a paper's abstract to train text classification models,
then apply those models to support the selection of other relevant papers. Since it is easy to implement
and performs reliably well, abstract-based methods are widely used in many domains~\citep{yu2018finding,miwa2014reducing,wallace2010semi,cormack2014evaluation}.

\subsection{Abstract-based Methods}
In this paper, we focus on   \emph{abstract-based} methods since:
\begin{itemize}
\item They are  easy to implement---the only extra cost is some negligible training time, compared to search query based methods which require human judgments on keywords selection and reference based methods which cost time and effort in extracting reference information.
\item Several studies show that the
performance of abstract-based methods is better than other approaches~\citep{task2,roegiest2015trec}. 
\end{itemize}
Note that abstract-based methods do not restrict the reviewers from making decisions based on full-text when it is hard to tell whether one paper should be included or not based on abstract. However, only abstracts are used for training and prediction, which makes data collection much easier. Abstract-based
 methods utilize different types of machine learning algorithms :

\begin{enumerate}

\item
\emph{Supervised learning}: which trains on labeled papers (that humans have already labeled as relevant/non-relevant) before classifying the remaining unlabeled papers automatically~\citep{cohen2006reducing,adeva2014automatic}. One problem with supervised learning methods is that they need a sufficiently large labeled training set to operate and it is extremely inefficient to collect such labeled data via random sampling. For this reason, supervised learners are often combined with active learning (see below) to reduce the cost of data collection.  
\item
\emph{Unsupervised learning}:   techniques like visual text mining (VTM) can be applied to facilitate the human labeling process~\citep{malheiros2007visual,felizardo2010approach}. In practice the cost reductions associated with (say) visual text mining is not as significant as that of \emph{active learning} methods due to not utilizing any labeling information~\citep{roegiest2015trec,task2}.
\item
\emph{Semi-supervised learning}: are similar to \emph{supervised learning} but utilizes unlabeled data to guide the training process. Different from active learning, it does not actively query new labels from unlabeled data. Instead, it tries to make use of the unlabeled data to improve the training result of a supervised learning~\citep{Chapelle2017Introduction}. Semi-supervised methods
have been proved to be more effective than supervised learning methods but still suffers from the same problem (the need for relatively large labeled training sets--\citet{liu2016comparative} reported a less than 95\% recall with 30\% data labeled).
\item
\emph{Active learning}:  In this approach, human reviewers read a few papers and classify each one as relevant or non-relevant. Machine learners then use this feedback to learn their models incrementally. These models are then used to sort the stream of papers such that humans read the most informative ones first. 
\end{enumerate}

\begin{algorithm}[!t]
\scriptsize
\SetKwInOut{Input}{Input}
\SetKwInOut{Output}{Output}
\SetKwInOut{Parameter}{Parameter}
\SetKwRepeat{Do}{do}{while}
\Input{$E$, set of all candidate papers\\$R$, set of ground truth relevant papers}
\Output{$L_R$, set of included papers}
\BlankLine

$L\leftarrow \emptyset$\; $L_R\leftarrow \emptyset$\; $\neg L\leftarrow E$\; 

\BlankLine
\tcp{Keep reviewing until stopping rule satisfied}
\While{$|L_R| < 0.95|R|$}{
    \tcp{Start training or not}
    \eIf{$|L_R| \geq 1$}{
        $CL\leftarrow Train(L)$\;
        \tcp{Query next}
        $x\leftarrow Query(CL,\neg L,L_R)$\;
    }{
        \tcp{Random Sampling}
        $x\leftarrow Random(\neg L)$\;
    }
    \tcp{Simulate review}
    $L_R,L\leftarrow Include(x,R,L_R,L)$\;
    $\neg L\leftarrow E \setminus L$\;
}
\Return{$L_R$}\;
\BlankLine
\Fn{Train($L$)}{
    \tcp{Train linear SVM with Weighting}
    $CL\leftarrow SVM(L,kernel=linear,class\_weight=balanced)$\;
    \If{$L_R\ge 30$}{
        \tcp{Aggressive undersampling}
        $L_I\leftarrow L\setminus L_R$\;
        $tmp\leftarrow L_I[argsort(CL.decision\_function(L_I))[:|L_R|]]$\;
        $CL\leftarrow SVM(L_R \cup tmp,kernel=linear)$\;
    }
    \Return{$CL$}\;
}
\BlankLine
\Fn{Query($CL,\neg L,L_R$)}{
    \eIf{$L_R< 10$}{
        \tcp{Uncertainty Sampling}
        $x\leftarrow argsort(abs(CL.decision\_function(\neg L)))[0]$\;
    }{
        \tcp{Certainty Sampling}
        $x\leftarrow argsort(CL.decision\_function(\neg L))[-1]$\;
    }
    \Return{$x$}\;
}
\BlankLine
\Fn{Include($x,R,L_R,L$)}{
    $L\leftarrow L \cup x$\;
    \If{$x\in R$}{
        $L_R\leftarrow L_R \cup x$\;
    }
    \Return{$L_R$, L}\;
}
\caption{Psuedo Code for FASTREAD~\citep{yu2018finding}}\label{alg:alg1}
\end{algorithm}

\subsection{FASTREAD}
\label{sect: FASTREAD}
\begin{figure}[!t]
    \centering
    \includegraphics[width=\linewidth]{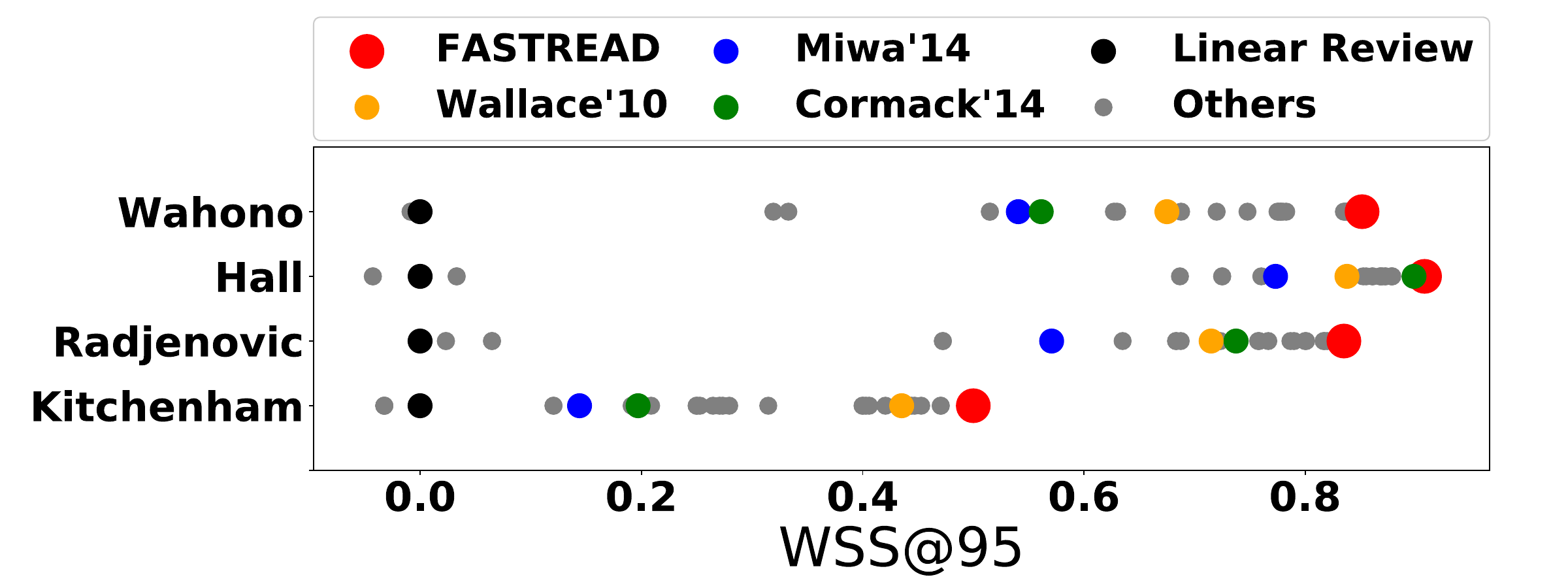}
    \caption{Median results (over 30 runs) of 32 active learning algorithms and manual review (linear review) across the four datasets of Table~\ref{tab: stats}. WSS@95 (higher the better) represents the effort (the number of papers need to be reviewed) saved when finding 95\% of the relevant papers. 
}
    \label{fig: Core}
\end{figure}

Our prior work in this area leads to the
FASTREAD abstract-based active learner~\citep{yu2018finding} shown in Algorithm~\ref{alg:alg1}.
FASTREAD employed the following   active learning strategies:
\begin{itemize}
\item Input tens of thousands of papers selected by (e.g.) a query to Google Scholar using techniques such as query expansion~\citep{Umemoto2016ScentBar}. 
\item
In the {\em initial random reasoning}   phase (shown in line 9 of
Algorithm~\ref{alg:alg1}), 
\begin{itemize}
\item
Allow humans to skim through the   papers manually 
until they have found $|L_R|\ge 1$ relevant paper (along with $|L_I|$ non-relevant ones). Typically, dozens of papers need to be examined to find the first relevant paper.
\end{itemize}
\item
After finding the first relevant paper, transit to the {\em reflective
reasoning} phase (shown in line 6 to 7 of
Algorithm~\ref{alg:alg1}):
\begin{itemize}
\item
Train an SVM model on the $L$ examples. When $|L_R|\ge 30$, balance the data via
{\em aggressive under-sampling}; i.e., reject all the non-relevant examples
{\em except} the $|L_R|$ non-relevant examples
furthest to the decision plane and on the non-relevant side.
\item
Use this model to decide what paper humans should read next.
Specifically, find the paper with the highest uncertainty (uncertainty sampling) when $|L_R|<30$ (shown in line 25 of Algorithm~\ref{alg:alg1}) and find the paper
with  highest probability to be relevant (certainty sampling) when $|L_R|\ge 30$ (shown in line 27 of
Algorithm~\ref{alg:alg1}). 
\item
Iterate the {\em reflective
reasoning} phase until more than 95\% of the relevant papers have been found ($|L_R|\ge 0.95|R|$).
\end{itemize}
\end{itemize}
FASTREAD was designed
after testing and comparing 32 different active
learners generated from three prior state of the art methods from the medical and legal literature---Wallace'10~\citep{wallace2010semi}, Miwa'14~\citep{miwa2014reducing}, and Cormack'14~\citep{cormack2014evaluation}. 
\begin{itemize}
\item
When compared to manual reading methods (linear review), FASTREAD dramatically reduces the effort associated with studies like those in Table~\ref{tab: stats}. With the help of FASTREAD, an order of magnitude fewer papers are required to be reviewed to find 95\% of the relevant ones.
\item
When compared to current state of the art automatic methods---Wallace'10~\citep{wallace2010semi}, Miwa'14~\citep{miwa2014reducing}, and Cormack'14~\citep{cormack2014evaluation}, FASTREAD reviews up to 50\% fewer papers while finding the same number of relevant papers.
\item
When compared to all the 32 methods used to design FASTREAD,
it performed remarkably better (see Figure~\ref{fig: Core}).
\end{itemize}

\subsection{Limitations with FASTREAD}
\label{sect: limit}
While  successful in some aspects,
FASTREAD suffers from three significant limitations.

\begin{figure}[!t]
    \centering
    \includegraphics[width=1\linewidth]{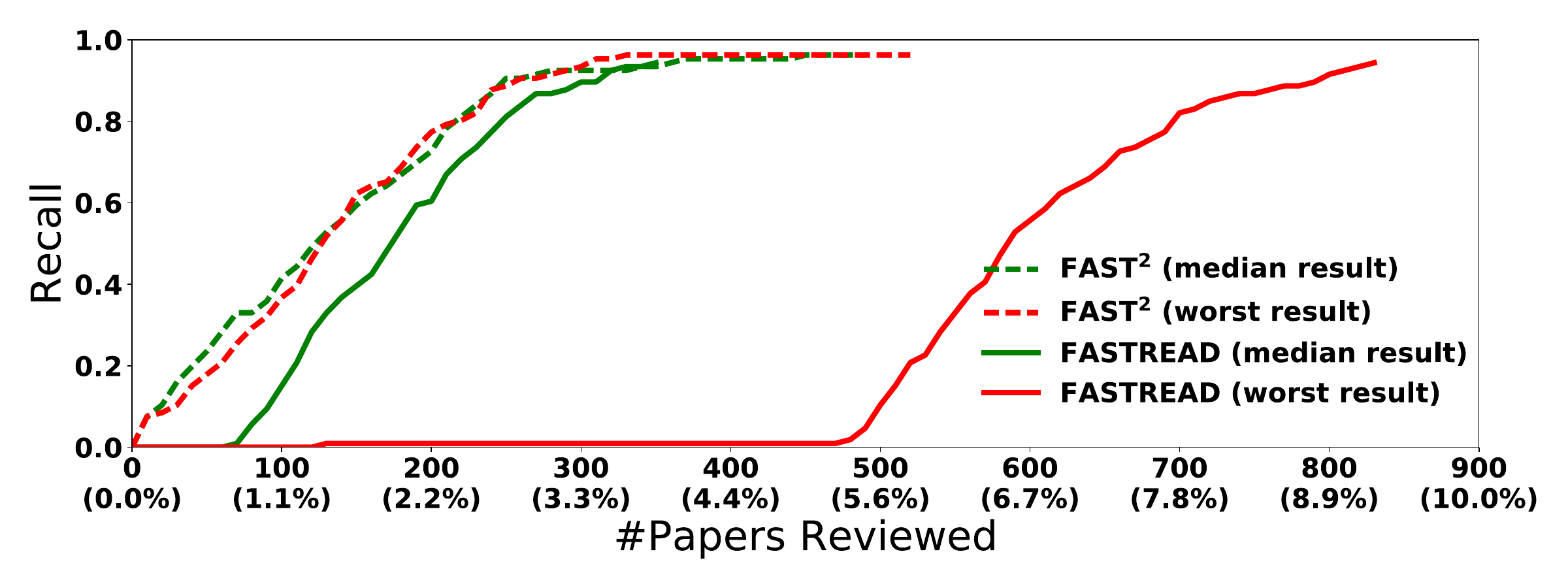}
    \caption{The performance of active learners can be extremely variable, depending on the initial choice of examples.
    This figure shows results from FASTREAD and FAST$^2$ across 30 random selections of initial papers. Note that in the worst case, it takes FASTREAD three times as long to find relevant papers (compared to the median case).
    Note also that the FAST$^2$ method proposed in this paper is far less susceptible to the adverse effects
    of poor initial example selection.}
    \label{fig: percentile}
\end{figure}

Firstly, FASTREAD utilizes {\em no domain knowledge} and knows nothing about its retrieval target before human oracles are provided.
Hence, the order of the papers explored in the initial reasoning phase is pure guesswork. This can lead to highly undesirable results, as shown in  Figure~\ref{fig: percentile}:
\begin{itemize}
 \item
 In that figure, the y-axis shows what percentage of the relevant papers were found (the data here comes from the Hall dataset in Table~\ref{tab: stats}).
 \item 
 The lines on this figure show the median (green) and worst-case (red) results in 30 runs,  we varied,
 at random, the papers used during initial reasoning.
 \item
 The dotted and solid lines show the performance of FASTREAD and FAST$^2$ respectively.
 \end{itemize}
Observe from the curves in Figure~\ref{fig: percentile}:
\begin{itemize}
\item
The solid curves denote FASTREAD's performance, and the distance between  FASTREAD's median and worst case is a factor of three.   This means
  the random selections made during the initial phase
 can have {\em runaway effects};
 i.e., an undesirable effect 
 of greatly increasing the effort associated with literature
 reviews. 
 \item
 On the other hand, as shown by the dashed curves of  Figure~\ref{fig: percentile},  FAST$^2$ has little variance and does not suffer from that {\em runaway effects}. As shown later
in this paper, FAST$^2$ achieves this via requesting minimal amounts of domain knowledge.
\end{itemize}
Secondly, using FASTREAD
{\em reviewers do not know when to stop reviewing}.
FASTREAD has no estimator that users can query
to ask ``is it worth reading any more papers?''. 
 Without such   knowledge  of  how  many  relevant  papers  that  are  left  to  be found, researchers might either:
 \begin{itemize}
 \item
 Stop too early, thus missing many relevant papers;
 \item Or stop too late, causing unnecessary effort even when there are no more relevant papers to find.
 \end{itemize}
The dashed lines in Figure~\ref{fig: percentile} show how FAST$^2$ stops near 0.95 recall without knowing the actual number of relevant papers to be found ($|R|$).  Section~\ref{sect: When to Stop} provides details about how FAST$^2$ achieves this.

\begin{figure}[!tbh]
    \centering
    \includegraphics[width=1\linewidth]{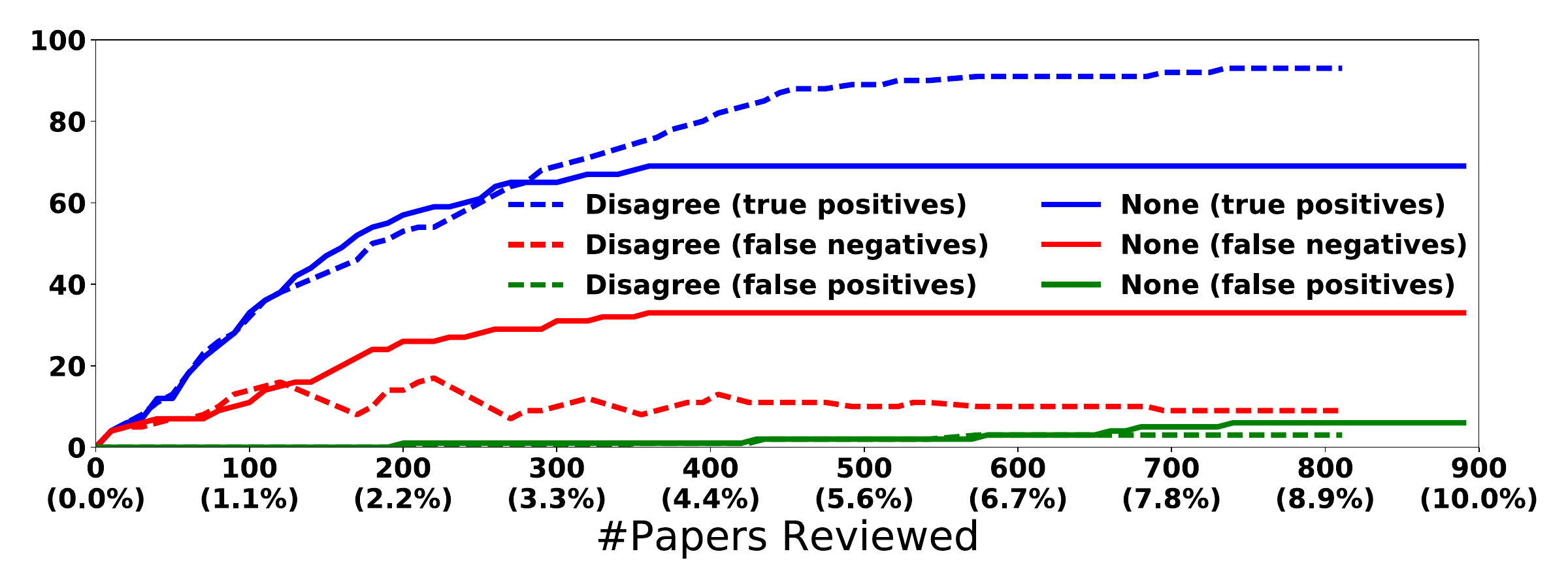}
    \caption{Simulation results for a user who can achieve 70\% precision and 70\% recall when reviewing all papers in Hall dataset. Blue curves show true positives (the number of correctly retrieved relevant papers); red curves show false negatives (the number of relevant papers but labeled as non-relevant); green curves show false positives (the number of non-relevant papers but labeled as relevant). Solid lines indicate the performance without error correction while dashed lines show improved performance with the error correction algorithm called \emph{Disagree}.}
    \label{fig: error}
\end{figure}

Thirdly, FASTREAD assumes
{\em reviewers are infallible};
i.e., when labeling papers
as relevant or non-relevant, they 
never make mistakes.
This seems a very optimistic assumption.
In practice, human errors are inevitable so any intelligent reading assistant should have some mechanism for handling those errors. Figure~\ref{fig: error} simulates a literature review on Hall dataset with a fallible human reviewer whose precision and recall are both 70\% (i.e., when asking the reviewer to label all papers, 70\% recall, and 70\% precision can be achieved). This is a reasonable assumption according to \citet{Cormack2017Navigating}. Solid lines show the performance without any error correction while dashed lines show the improved results (with increasing true positives and decreasing false negatives and false positives) with error correction method in FAST$^2$. Details of the error correction method in FAST$^2$ will be provided in Section~\ref{sect: How to correct}.


\section{How to Start}
\label{sect: How to Start}

The rest of this paper explores ways to resolve the limitations of FASTREAD.
In this first  section,
we drop the assumption of \emph{no  domain knowledge} to answer the following research question:

{\bf RQ1: How to start?} How to better utilize external domain knowledge to boost the review process and avoid the runaway effects illustrated in Figure~\ref{fig: percentile}?

\subsection{Related Work}
\label{sect: How to Start: Related Works}

Applying domain knowledge to guide the initial selection of training data, instead of random sampling, can avoid efforts being wasted on non-relevant examples and prevents \emph{runaway} results. However, no matter what form of domain knowledge applied, a common issue will be raised when random sampling is replaced; i.e., how to collect unbiased non-relevant examples. Without random sampling, the training data collected by domain knowledge guided review, uncertainty sampling, or certainty sampling are all biased and lead to a deteriorated performance when used for training. 

Our solution to this problem is based on   {\em presumptive non-relevant examples} (shortened as PRESUME in the rest of this paper)  proposed in the legal literature by \citet{cormack2015autonomy}. Each time before training, PRESUME samples randomly from the unlabeled examples and presumes the sampled examples to be non-relevant in training. The rationale behind this technique is that given the low prevalence of relevant examples, it is likely that most of the presumed ones are non-relevant. 

With the help of PRESUME, \citet{cormack2015autonomy} investigated two different ways to utilize different forms of domain knowledge for a better collection of initial training data:
\begin{itemize}
\item
{Auto-BM25:} keyword search and rank the unlabeled examples with their corresponding BM25 scores, which is calculated based on tf-idf~\citep{robertson2009probabilistic}. Then label the example with highest BM25 score as relevant and start training. Given a search query $Q=\{q_1,q_2,\dots,q_n\}$, the BM25 score is calculated as
\begin{equation}\label{eq:bm25}
\begin{aligned}
&BM25(i,Q)\\
=&\sum_{j=1}^{|Q|} IDF(q_j)\frac{f(q_j,i)}{f(q_j,i)+k_1(1-b+b\frac{l_i}{avgdl})}
\end{aligned}
\end{equation}
where $f(q_j,i)$ is term $q_j$'s term frequency in document $i$, $l_i$ is the length of document $i$, and $avgdl$ is the average document length in the collection $C$. $k_1$ and $b$ are free variables, here, we choose $K_1=1.5$, $b=0.75$. $IDF(q_j)$ is the inverse document frequency weight of term $q_j$ and it is computed as:
$$IDF(q_j)=log\frac{|E|-n(q_j)+0.5}{n(q_j)+0.5}$$ where $n(q_j)$ is the number of documents containing term $q_j$.
\item
{Auto-Syn:} a synthetic document is created, with domain knowledge from the reviewer, to act as the first relevant example and train the classifier directly.
\end{itemize}

\subsection{Method}
\label{sect: How to start: Method}

While the above methods from \citet{cormack2015autonomy} were shown to be useful in their test domain,
they have certain limitations.
Both methods rely on domain knowledge from reviewers to generate one seed relevant example and then use it to initiate the active learning process. One problem with such techniques is that the quality of the seed example, which is decided by the reviewers' expertise, can affect the performance of the active learner. More efforts are required to find new relevant examples if the seed example is not representative enough. 

To address this concern, in our work we modified {Auto-BM25} and created a new method called {Rank-BM25}:
\begin{itemize}
\item Just as with  {Auto-BM25}, {Rank-BM25} starts with querying a set of keywords and rank the unlabeled examples based on their BM25 scores.
\item
Rank-BM25 then
 asks a reviewer to review $N=10$ examples with descending order of their BM25 scores.
\item
 If $|L_R| \ge 1$ relevant examples found, start active learning; otherwise Rank-BM25 tries a different keyword set. 
\end{itemize}

\subsection{Experiments}
\label{sect: How to Start: Experiments}

In this subsection, we design experiments to answer \textbf{RQ1} by comparing different ways of utilizing external domain knowledge using the methods
described above.

\subsubsection{Datasets}
\label{sect: datasets}

Previously, we have created three datasets by reverse-engineering existing SLRs~\citep{yu2018finding}. The three datasets are named after the authors of their original publication source---Wahono dataset from \citet{wahono2015systematic}, Hall dataset from \citet{hall2012systematic}, and Radjenovi{\'c} dataset from \citet{radjenovic2013software}.  Apart from the three created datasets, one dataset (Kitchenham) is provided directly by the author of \citet{kitchenham2013systematic}. Statistics of the four datasets are shown in Table~\ref{tab: stats}. In this paper, we use the same four datasets to evaluate different techniques. All the above datasets are available online~\citep{fastread}.

It is appropriate to ask why we used these four datasets and not four others?
The answer is that, in the domain of automated support for reading the SE
literature, there is very little data available for experimentation.
We had to build three of the data sets from the last paragraph---a process
that took weeks of work. Only after extensively advertising our technique did we receive offers of other data from other researchers (and that
lead to the four datasets contributed by Kitchenham).

\subsubsection{Performance Metrics}
\label{sect: How to Start: Performance Metrics}

In order to compare with FASTREAD, the same performance metrics are applied here:
\begin{itemize}
\item
X95 = $\min \{|L| \mid |L_R|\geq0.95 |R|\}$.
\item
WSS@95 = $0.95-\text{X95}/|P|$.
\end{itemize}
X95 stands for the number of papers need to be reviewed in order to get $0.95$ recall, while WSS@95 represents the work saved, comparing to random sampling, when reaching $0.95$ recall. The reason behind 
$0.95$ \textbf{recall} is that a) $1.00$ \textbf{recall} can never be guaranteed by any text mining method unless all the candidate papers are reviewed; b) $0.95$ \textbf{recall} is usually considered acceptable in evidence-based medicine~\citep{cohen2011performance,cohen2006reducing,o2015using} despite the fact that there might still be relevant papers missing~\citep{shemilt2016use}.

As recommended by Mittas \& Angelis in their 2013 IEEE TSE paper~\citep{mittas2013ranking}, Scott-Knott analysis was applied to cluster and rank the performance (X95) of each treatment. It clusters treatments with little difference in performance together and ranks each cluster with the median performances~\citep{scott1974cluster}. As suggested in the authors' prior work~\citep{yu2018finding}, nonparametric hypothesis tests are applied to handle the non-normal distribution. Specifically, Scott-Knott decided two methods are not of little difference if both bootstrapping~\citep{efron1982jackknife}, and an effect size test~\citep{cliff1993dominance} agreed that the difference is statistically significant ($99\%$
confidence) and not a negligible effect (Cliff's Delta $\ge$ 0.147). 

\subsubsection{Treatments}
\label{sect: How to Start: Treatments}

As for \textbf{RQ1: ``How to start?''} four different treatments are tested, including the random sampling tactic from FASTREAD~\citep{yu2018finding} as a baseline, Auto-BM25 and Auto-Syn from \citet{cormack2015autonomy}, and Rank-BM25 created in this paper. To compare performance and variance of different treatments, each one is tested on all four datasets with 30 random seeds. For each treatment, the corresponding domain knowledge imported is as following:
\begin{itemize}
\item
Rank-BM25: ``Topic'' column in Table~\ref{tab: stats} as the search query $Q$.
\item
Auto-BM25: ``Topic'' column in Table~\ref{tab: stats} as the search query $Q$.
\item
Auto-Syn:  ``Original Title'' column in Table~\ref{tab: stats} as the synthetic document.
\item
FASTREAD: random sampling until the first relevant paper is found, same as FASTREAD.
\end{itemize}
Note that for all treatments, PRESUME is required to keep the training data unbiased.

\begin{table*}[!t]
\caption{Testing tactics for ``how to start''}
\label{tab: start}
\begin{center}
\begin{threeparttable}
\small
\setlength\tabcolsep{15pt}
\begin{tabular}{ l|c|l|c|c|c|c }
 \multicolumn{3}{c}{} & \multicolumn{2}{c|}{X95} & \multicolumn{2}{c}{WSS@95} \\
\cline{4-7}
 Dataset & Rank & Treatment & median & IQR & median & IQR \\
\hline
\multirow{4}{*}{Wahono} & 1 & Rank-BM25 & 630 & 48 & 0.86 & 0.01 \\
& 2 & FASTREAD &  685 & 225 & 0.85 & 0.03 \\
& 2 & Auto-Syn  & 705 & 60 & 0.85 & 0.01 \\
& 3 & Auto-BM25  & 785 & 78 & 0.84 & 0.01 \\
\hline
\multirow{4}{*}{Hall} & 1 & Rank-BM25 & 290 & 10 & 0.92 & 0.00 \\
& 2 & Auto-Syn & 320 & 30 & 0.91 & 0.00 \\
& 3 & Auto-BM25 & 345 & 30 & 0.91 & 0.00  \\
& 3 & FASTREAD & 345 & 125 & 0.91 & 0.01 \\
\hline
 \multirow{4}{*}{Radjenovi{\'c}} & 1 & Auto-Syn & 515 & 122 & 0.86 & 0.02 \\
 & 2 & Rank-BM25 & 615 & 85 & 0.85 & 0.01 \\
 & 3 & FASTREAD & 700 & 208 & 0.83 & 0.03 \\
 & 4 & Auto-BM25 & 800 & 135 & 0.82 & 0.02 \\
\hline
\multirow{4}{*}{Kitchenham} & 1 & Auto-BM25 & 510 & 30 & 0.65 & 0.02 \\
& 1 & Auto-Syn & 520 & 40 & 0.64 & 0.02 \\
& 1 & Rank-BM25 & 525 & 60 & 0.64 & 0.03 \\
& 2 & FASTREAD & 630 & 130 & 0.58 & 0.07  \\
\hline
\end{tabular}
\begin{tablenotes}\small
Each experiment is repeated 30 times. Only medians and IQR (75-25th percentile, smaller IQR means less variances) are shown in this table. On each dataset, different starting tactics are compared by their X95 and WSS@95 scores (X95 smaller the better, WSS@95 larger the better). Scott-Knott analysis is applied to rank different treatments in ``Rank'' column. Treatments ranked the same by Scott-Knott analysis are considered to have similar performance while treatments ranked differently are significantly different in performance.
\end{tablenotes}
\end{threeparttable}
\end{center}
\end{table*}

\subsection{Results}
\label{sect: How to Start: Results}

Table~\ref{tab: start} shows the results of different starting tactics on four SE SLR datasets. Medians and IQRs (75-25th percentile, smaller IQR means less variances) are shown in this table. On each dataset, different treatments are compared by their X95 and WSS@95 scores (X95 smaller the better, WSS@95 larger the better). Scott-Knott analysis is applied to rank different treatments in ``Rank'' column. Treatments ranked the same by Scott-Knott analysis are considered to have similar performance while treatments ranked differently are significantly different in performance. With the results in Table~\ref{tab: start}, we reach the following conclusions.

{Rank-BM25} is recommended as the overall best tactic. It is the
only tactic which consistently performs better than the baseline tactic {FASTREAD} (top rank in 3 out of 4 datasets, second rank in the other dataset). 
While this tactic requires two or three keywords to initiate, they can be easily found in the original search string.

{Auto-BM25} is considered as a depreciated version of {Rank-BM25} since it never ranks better than {Rank-BM25} (worse in three and similarly in one dataset).

{Auto-Syn} is not recommended.  It requires human experts to synthesize example relevant text, which is a difficult task. The quality of the synthesized example may affect the performance a lot, thus leads to unstable results (it performs the best in two datasets but also performs worse, though not significantly, than the baseline in one dataset). 

{FASTREAD} is not recommended since domain knowledge as trivial as two or three keywords can save 10-20\% more review efforts and reduce the IQRs greatly (e.g., IQRs of Rank-BM25 are $30-50\%$ of those of Random). This indicates better robustness of the active learner and runaway results (where some readings take far longer than others due to random selection of initial papers)
are far less likely when applying domain knowledge to start the review. 
 
Hence we say:

\begin{RQ}{Answer to RQ1 ``How to Start?''}
    Overall, we suggest {Rank-BM25} as the most effective starting tactic for initiating active learning. By applying such tactic, the review effort can be reduced by 10-20\% by FASTREAD while preventing the \emph{runaway effects}.
\end{RQ}

\begin{algorithm}[!t]
\scriptsize
\SetKwInOut{Input}{Input}
\SetKwInOut{Output}{Output}
\SetKwInOut{Parameter}{Parameter}
\SetKwRepeat{Do}{do}{while}
\Input{$E$, set of all candidate papers\\$R$, set of ground truth relevant papers\\$Q$, search query for BM25}
\Output{$L_R$, set of included papers}
\BlankLine

$L\leftarrow \emptyset$\; $L_R\leftarrow \emptyset$\; $\neg L\leftarrow E$\; 

\BlankLine
\While{$|L_R| < 0.95|R|$}{
    \eIf{$|L_R| \geq 1$}{
        \tcp{Presumptive non-relevant examples}
        \colorbox{yellow!25}{$L_{pre}\leftarrow Presume(L,\neg L)$}\;
        \colorbox{yellow!25}{$CL\leftarrow Train(L \cup L_{pre})$}\;
        $x\leftarrow Query(CL,\neg L,L_R)$\;
    }{
        \tcp{BM25 ranking with keywords $Q$}
        \colorbox{yellow!25}{$x\leftarrow argsort(BM25(\neg L,Q))[0]$}\;
    }
    $L_R,L\leftarrow Include(x,R,L_R,L)$\;
    $\neg L\leftarrow E \setminus L$\;
}
\Return{$L_R$}\;

\BlankLine
\Fn{Presume($L,\neg L$)}{
    \tcp{Randomly sample $|L|$ points from $\neg L$}
    \Return $Random(\neg L,|L|)$\;
}

\caption{Psuedo Code for Rank-BM25 on FAST$^2$}\label{alg:alg2}
\end{algorithm}

The updated algorithm with {Rank-BM25} is shown in Algorithm~\ref{alg:alg2}. BM25 score is calculated as \eqref{eq:bm25}. Rows colored in
yellow
are the differences from Algorithm~\ref{alg:alg1}. Functions already described in Algorithm~\ref{alg:alg1} are omitted.

\section{When to Stop}
\label{sect: When to Stop}

This section drops the assumption of \emph{reviewer knows when to stop reviewing} on top of Algorithm~\ref{alg:alg2} and answers the following research question:

{\bf RQ2: When to stop?} Without knowing the actual number of relevant papers in the candidate set, how to decide when to stop reviewing so that 1) most relevant papers have been retrieved, 2) not too much review effort is wasted. 

\subsection{Related Work}
\label{sect: When to Stop: Related Works}

When to stop is a critical problem while applying active learning on literature reviews. If the reading stops too early, it may end up with missing too many relevant papers. On the other hand, if the reading stops too late, review effort might be wasted since no new relevant papers can be found. 

Before this study,
other work~\citep{wallace2010semi,miwa2014reducing} usually focus on generating the best reading curve and do not discuss the stopping rule. 
That said, our reading of the literature is that there   exists three candidate state of the art stopping rules for selecting relevant papers in software engineering, electronic discovery, and evidence-based medicine respectively:
\begin{itemize}
\item
Ros'17: stop review after 50 non-relevant papers are found in succession~\citep{ros2017machine}.
\item
Cormack'16: the knee method~\citep{Cormack2016Engineering}. This method detects the inflection point $i$ of current recall curve, and compare the slopes before and after $i$. If $slope_{<i}/slope_{>i}$ is greater than a specific threshold $\rho$, the review should be stopped. For details about this knee method, please refer to \citet{Cormack2016Engineering}.
\item
Wallace'13: Apply an estimator to estimate the number of relevant papers $|R|$ and let the users decide when to stop by showing them how close they are to the estimated number~\citep{wallace2013active}.
\end{itemize}
We note that the Ros'17 and Cormack'16 methods are less flexible than the Wallace'13 method since they do not allow the user to choose what recall they want. Also note that another recently proposed stopping strategy by \citet{di2018study} requires the review to follow the order of BM25, therefore is not compared in this papers.

\subsection{Method}
\label{sect: When to Stop: Method}

According to \citet{cohen2011performance}, we want the reading to stop at 95\% recall. To achieve this, we must first know the total number of relevant papers $|R|$. Since this is unknowable in practice, the ``when to stop'' problem can be solved by building a good class probability estimator which can tell the reviewer a) what is the probability of being relevant for each unlabeled paper; b) how many relevant papers have not been found yet. 

The challenges for building such a class probability estimator include the class imbalance of training data~\citep{wallace2012class} and the bias incurred by sampling with active learning. Specifically, given that active learning provides a better chance to retrieve relevant papers in an early stage, the prevalence of relevant papers in the reviewed set $P_L=|L_R|/|L|$ is much larger than that in the candidate set $P=|R|/|E|$~\citep{wallace2013active}. 

Wallace'13 solved the above challenges by {sampling $\propto$ probabilities}---sample unlabeled examples with a probability of active learner's prediction. However, {sampling $\propto$ probabilities} is less efficient than the query strategy of FASTREAD (details will be presented in Section~\ref{sect: When to Stop: Results}). Therefore we design a new estimator {SEMI} which utilizes the same query strategy of FASTREAD and fits a semi-supervised logistic regressor utilizing not only the labeled data but also the unlabeled data. 

SEMI utilizes a recursive \emph{TemporaryLabel} technique. Each time the SVM model is retrained, SEMI assigns temporary labels to unlabeled data points and builds a logistic regression model on the temporary labeled data. It then uses the obtained regression model to predict on the unlabeled data and updates the temporary labels. This process is iterated until convergence. Algorithm~\ref{alg:alg3} shows how {SEMI} works in detail where the \emph{LogisticRegression} function is implemented with scikit-learn, and its regularization strength $C$ is calculated as $C=|R_E|/(|L|-|L_R|)$ to mitigate the imbalance in SVM classifier.

\begin{figure*}[!htb]
    \centering
    \setlength\tabcolsep{4pt}
    \begin{tabular}{l|cc}
    \setlength\tabcolsep{2pt}
     & \makecell{Effectiveness of different active learning strategies  } & \makecell{Predictive error, while querying   } \\
     
      & \makecell{  (higher values are better)} & \makecell{  (values nearer ``true'' are  better) } \\

    \hline
    \rotatebox[origin=l]{90}{\parbox[c]{2.8cm}{\centering \textbf{Wahono}}} &  \includegraphics[width=0.43\linewidth]{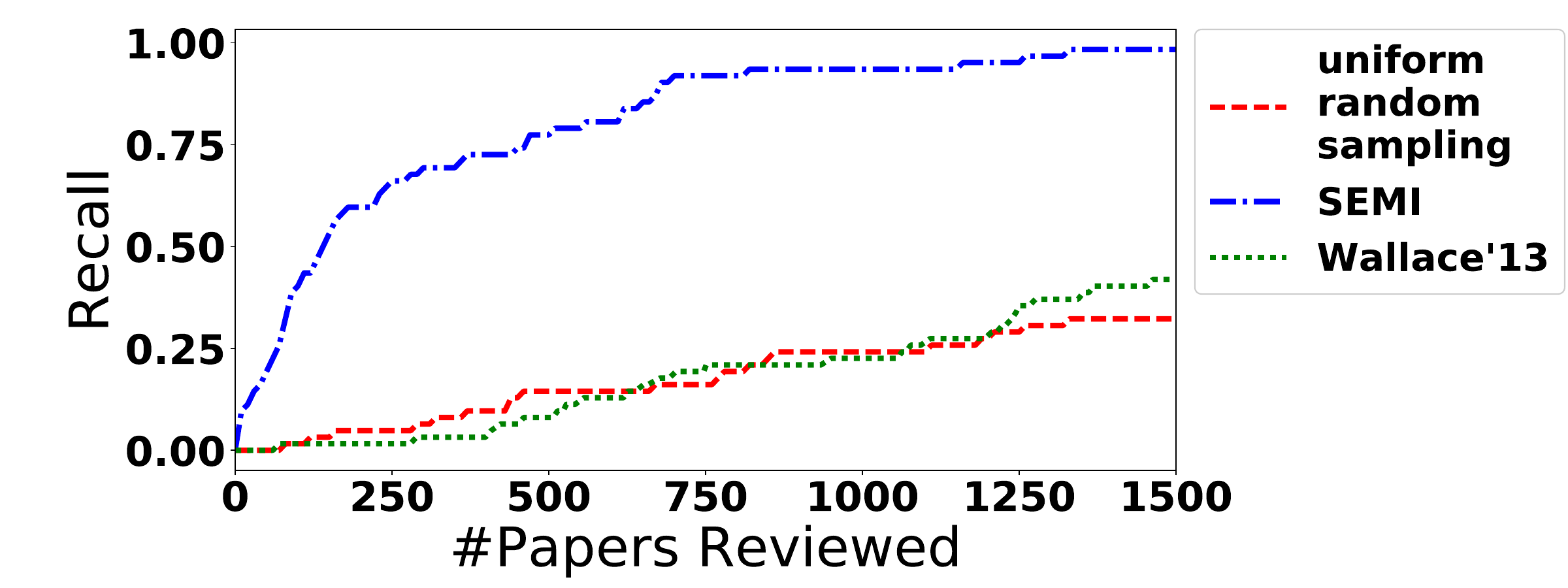} & \includegraphics[width=0.43\linewidth]{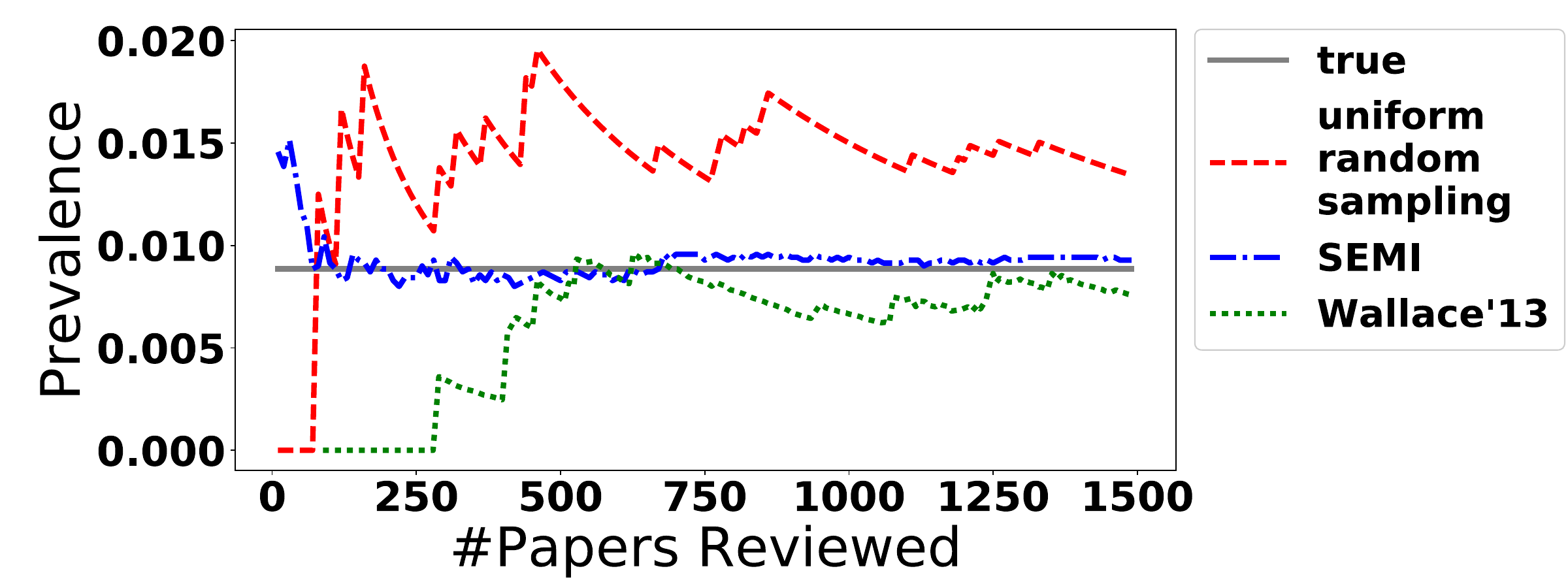} \\
    \rotatebox[origin=l]{90}{\parbox[c]{2.8cm}{\centering \textbf{Hall}}} & \includegraphics[width=0.43\linewidth]{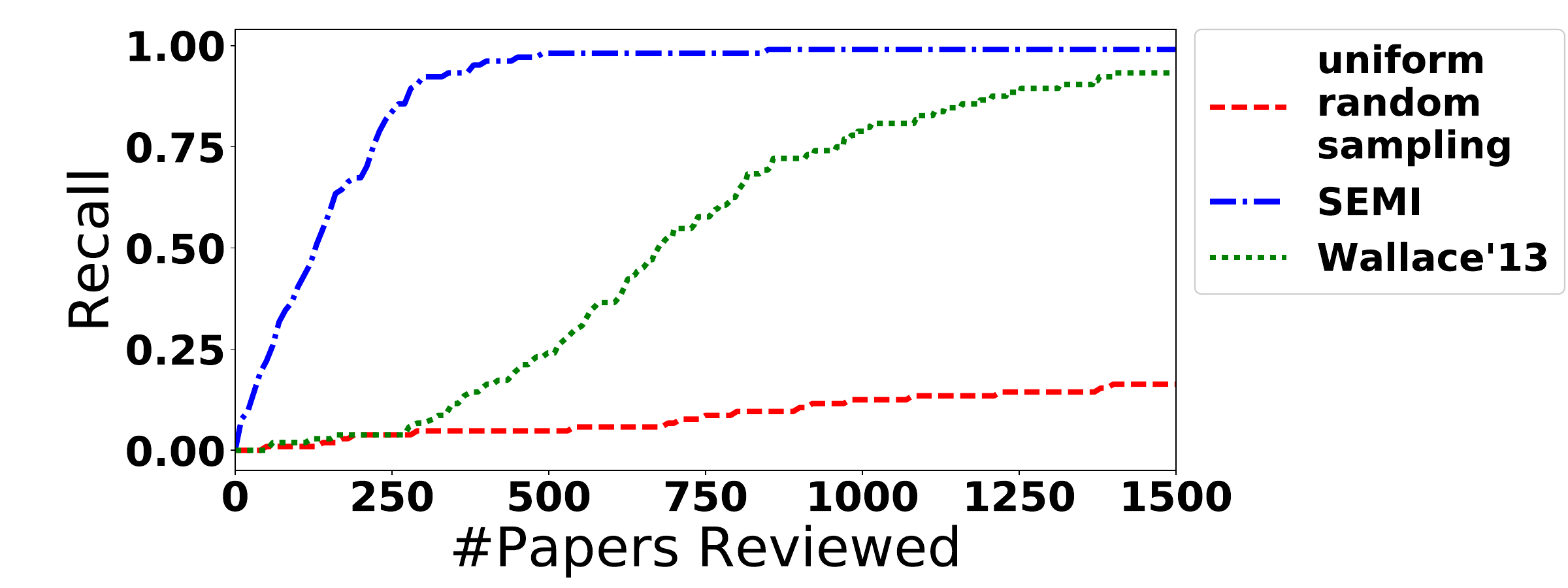} & \includegraphics[width=0.43\linewidth]{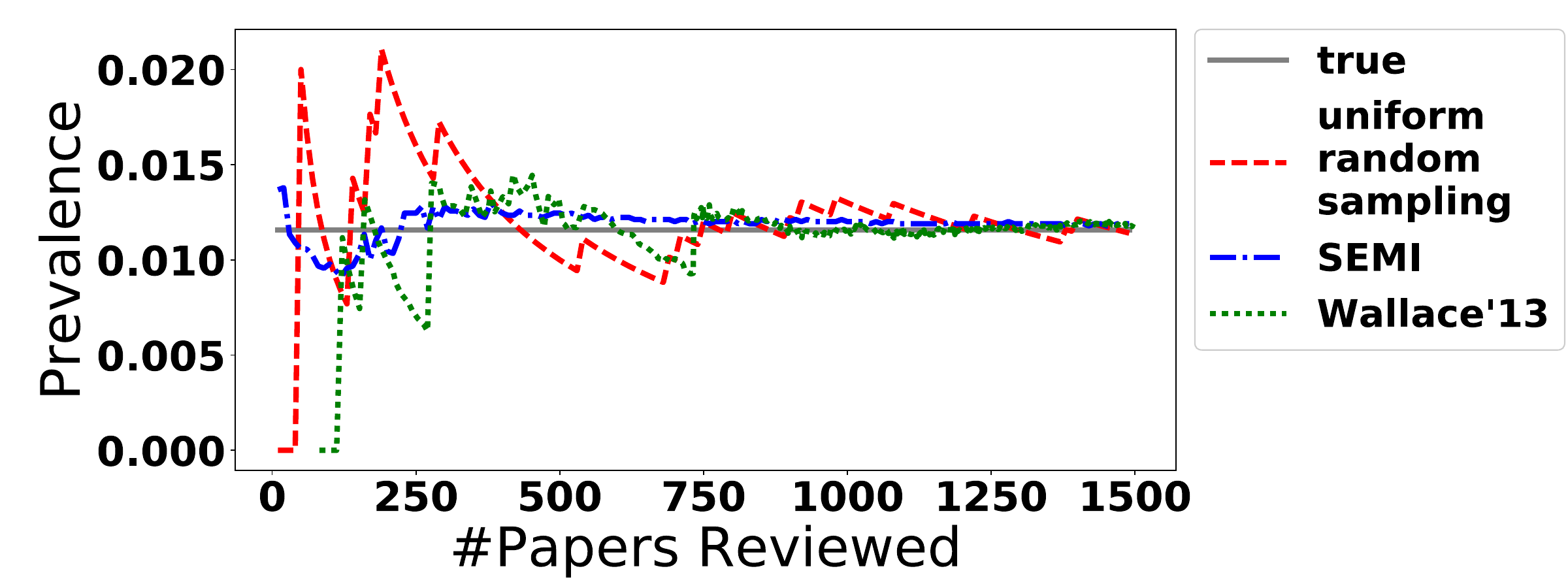}\\
    \rotatebox[origin=l]{90}{\parbox[c]{2.8cm}{\centering \textbf{Radjenovi{\'c}}}} & \includegraphics[width=0.43\linewidth]{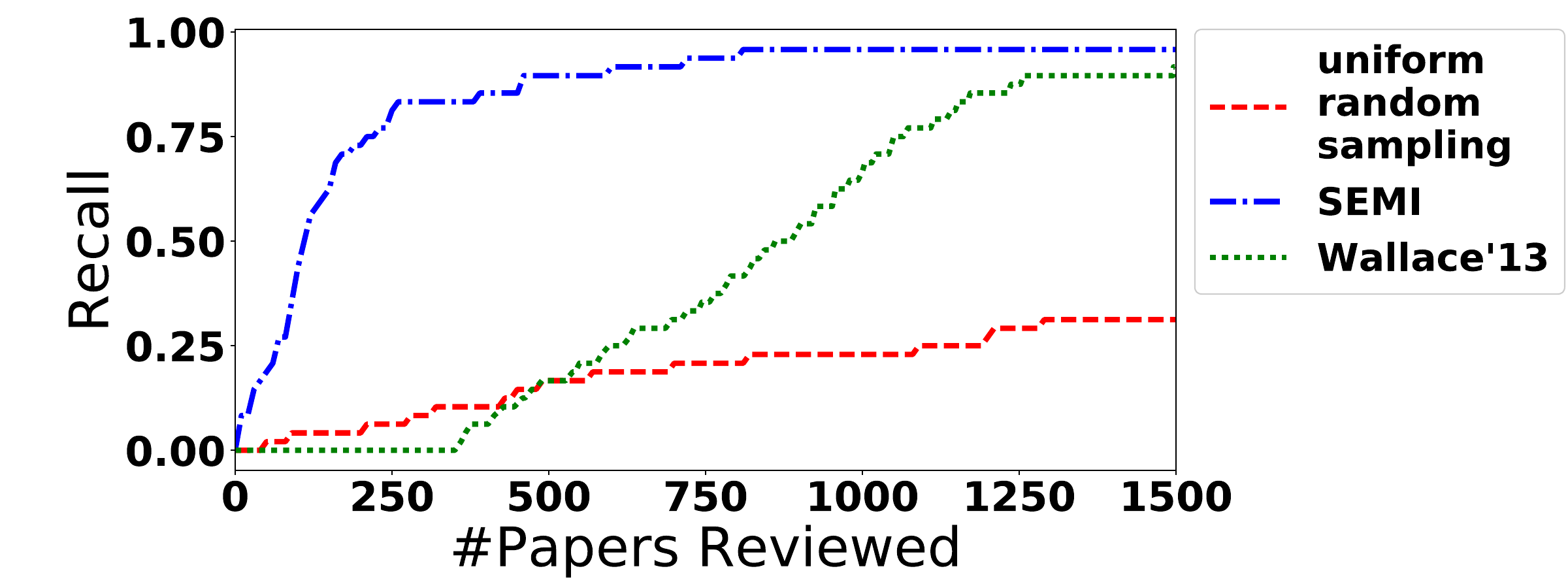} & \includegraphics[width=0.43\linewidth]{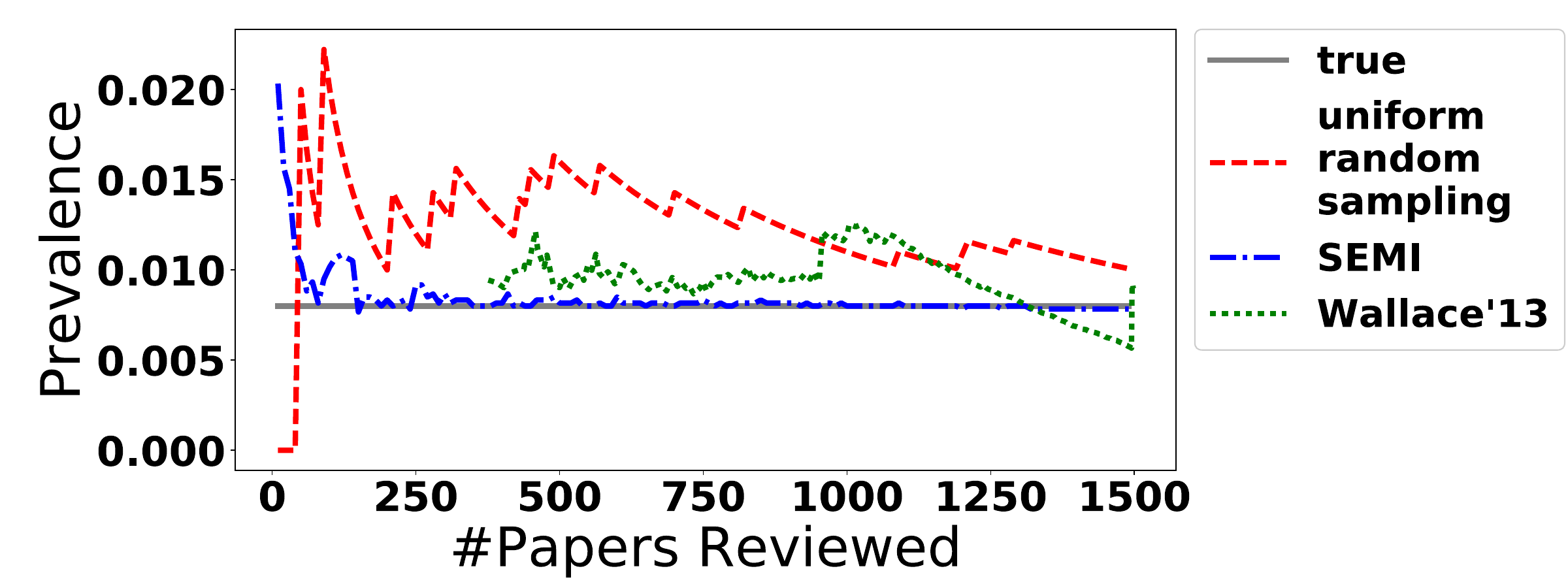}\\
    \rotatebox[origin=l]{90}{\parbox[c]{2.8cm}{\centering \textbf{Kitchenham}}} & \includegraphics[width=0.43\linewidth]{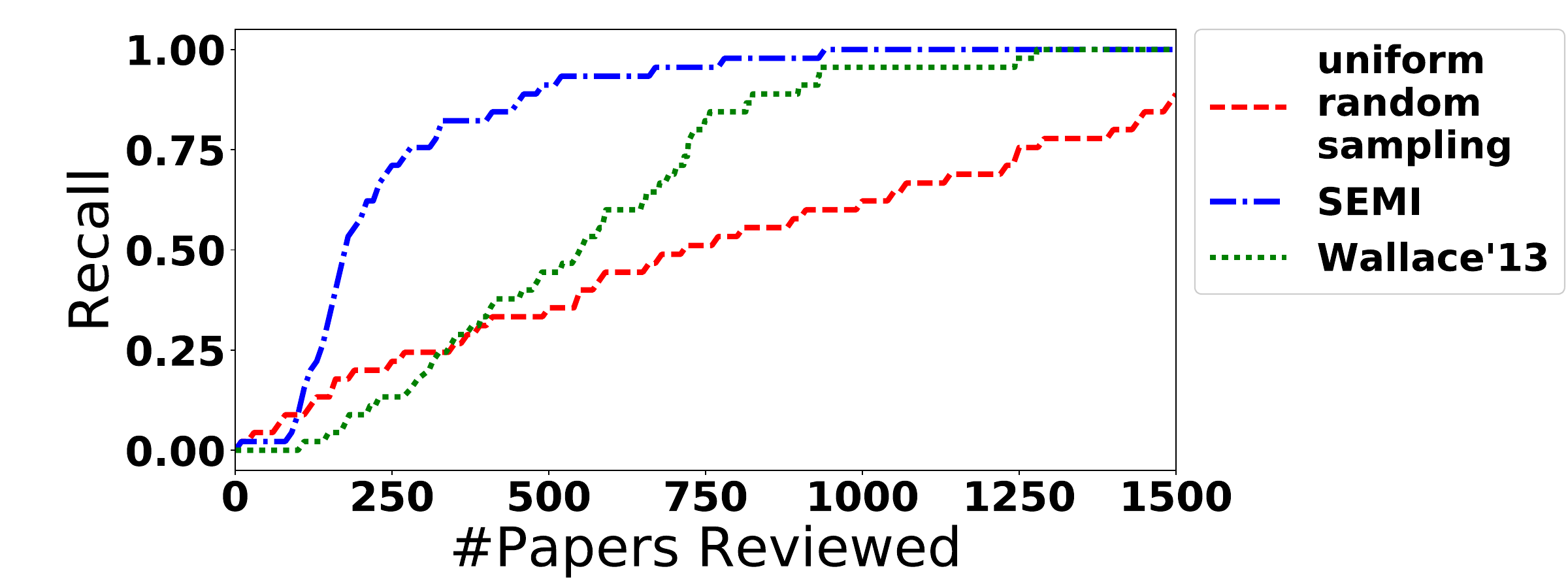} & \includegraphics[width=0.43\linewidth]{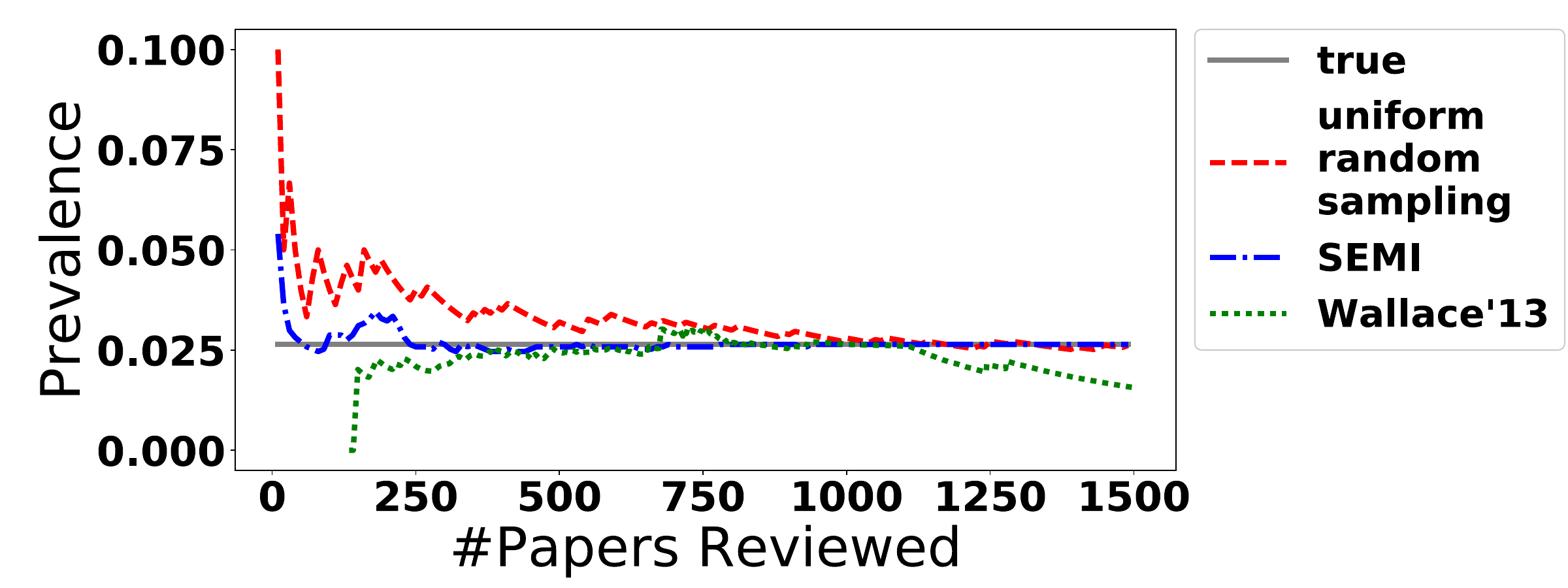}\\
    \end{tabular}
    \caption{Results for prevalence estimation ($|R_E|/|E|$) with each row on one different dataset. Estimation from uniform random sampling works as a baseline (see the red line).  SEMI (in blue) finds relevant papers much faster than uniform random sampling (in red) or Wallace'13 (in green). The second column demonstrates the effectiveness of different estimators where one estimator is more accurate than another if its estimation is closer to the true prevalence (in gray) with fewer papers reviewed. {SEMI} estimator (in blue) estimates more accurately than both the Wallace estimator (in green) and the estimation from uniform random sampling (in red).}
    \label{fig: estimate}
\end{figure*}

\subsection{Experiments}
\label{sect: When to Stop: Experiments}

In this subsection, we design experiments to answer \textbf{RQ2} by 1) comparing the estimator of SEMI with Wallace'13, 2) comparing the performance of applying SEMI as stopping rule with Ros'17 and Cormack'16 stopping rules.

Firstly, {SEMI} was compared with the Wallace'13 estimators for the task of estimating the prevalence of relevant examples ($|R|/|E|$). Results from uniform random sampling~\footnote{Uniform random sampling samples from unlabeled data randomly and queries the labels, it then estimates $|R|$ as $|R_E|=|E|\times|L_R|/|L|$. It was observed to estimate more accurately than the Wallace'13 estimator according to \citet{wallace2013active}.} is also provided as a baseline. 

We then used the SEMI estimator as an early stopping rule for literature reviews. Reviews were simulated on our four full datasets and stopped when $|L_R|\geq T_{rec}|R_E|$. $T_{rec}$ represents the desired final recall of the review and $T_{rec}=0.90$ for SEMI (90), $T_{rec}=0.95$ for SEMI (95). Note that $R_E$ was calculated each time the learner was retrained. Performances of this new stopping rule were analyzed by comparing with :
\begin{itemize}
\item
Ros'17: i.e., stopping when 50 non-relevant papers found in succession; 
\item
Cormack'16: The knee method with $\rho= 6$.
\end{itemize}
Note that besides $\rho= 6$, \citet{Cormack2016Engineering} have suggested an alternative $\rho= 156 - min(|L_R|, 150)$, especially for the cases of a relatively small dataset. However when we tested it, Cormack'16 stopping rule barely worked with $\rho= 156 - min(|L_R|, 150)$ and almost all candidate papers were reviewed. Therefore we choose $\rho= 6$ for the  Cormack'16 method.

\subsection{Results}
\label{sect: When to Stop: Results}

Before applying SEMI as a stopping rule, the accuracy of its estimation is first tested against that of the prior state of the art estimator---the Wallace'13 estimator. Figure~\ref{fig: estimate} shows the performances of different estimators. The first column in Figure~\ref{fig: estimate} demonstrates the effectiveness of different query strategies. Suggested by these results, SEMI (in blue) outperforms Wallace'13 (in green) and uniform random sampling (in red) since it achieves same recall with much fewer papers reviewed. 

The second column of Figure~\ref{fig: estimate} demonstrates the effectiveness of different estimators. The {SEMI} estimator (in blue) is not only more accurate than the Wallace'13 estimator (in green) but also the estimation from uniform random sampling (in red). SEMI is a better estimator than the state of the art Wallace'13 estimator since both its query strategy and its estimation are better. Moreover, in all our case studies, after reviewing 150 documents, the SEMI estimator becomes remarkably accurate. This means that, very early in a literature review,  researchers can adjust their review plans according to the number of remaining papers.

Then we apply SEMI as the stopping rule and compare it against other state of the art stopping rules. Based on the results shown in Table~\ref{tab: stopping}, we observe that 
Ros'17 usually stops too early and results in a low recall; Cormack'16 results in a large range of recalls which the user has no control of. On the other hand, we suggest SEMI as the best stopping rule since (1) it has the advantage of letting user to choose the target recall ($T_{rec}$) and (2) based on the results, it stops very close to the desired recall stably (with lower IQRs).

\begin{table}[!htb]
\caption{Comparison of different stopping rules}
\label{tab: stopping}
\begin{threeparttable}
\small
\setlength\tabcolsep{4pt}
\begin{tabular}{ l|l|c|c }
  \textbf{Dataset} & \textbf{Stopping rule} & \makecell{\textbf{Final Recall} \\$|L_R|/|R|$} & \makecell{\textbf{\#papers}\\ \textbf{reviewed} $|L|$}\\
  \hline
  \multirow{4}{*}{Wahono} & SEMI (90) & 94(0) & 750(48)\\
   & SEMI (95) & 95(0)  & 1180(67) \\
   & Ros'17 & 76(8)  & 370(108) \\
   & Cormack'16 & 94(11)   & 770(288)\\
  \hline
  \multirow{4}{*}{Hall} & SEMI (90) &  95(2)  & 315(20) \\
  & SEMI (95) & 98(0) & 500(30)  \\
  & Ros'17 & 97(0)  & 385(48) \\
  & Cormack'16 & 97(1)   & 390(38)\\
  \hline
  \multirow{4}{*}{Radjenovi{\'c}} & SEMI (90) & 92(0) & 540(58) \\
  & SEMI (95) & 94(0) & 760(37) \\
  & Ros'17 & 83(4)  & 320(75) \\
  & Cormack'16 & 88(6)   & 420(100)\\
  \hline
  \multirow{4}{*}{Kitchenham} & SEMI (90) & 89(2) & 400(20) \\
  & SEMI (95) & 91(2) & 510(20) \\
  & Ros'17 & 89(0)  & 420(50) \\
  & Cormack'16 & 96(2)   & 660(85)\\
  \hline
\end{tabular}
\begin{tablenotes}\small
Each experiment is repeated for 30 times. Results are shown as median(IQR) in percentage. As for recall, the higher the better; as for \#papers reviewed, the fewer the better.
\end{tablenotes}
\end{threeparttable}
\end{table}

\begin{algorithm}[!htp]
\scriptsize
\SetKwInOut{Input}{Input}
\SetKwInOut{Output}{Output}
\SetKwInOut{Parameter}{Parameter}
\SetKwRepeat{Do}{do}{while}
\Input{$E$, set of all candidate papers\\$R$, set of ground truth relevant papers\\$Q$, search query for BM25\\\colorbox{yellow!25}{$T_{rec}$, target recall (default=0.95)}}
\Output{$L_R$, set of included papers}
\BlankLine

$L\leftarrow \emptyset$\; $L_R\leftarrow \emptyset$\; $\neg L\leftarrow E$\; 
\colorbox{yellow!25}{$|R_E|\leftarrow \infty$}\;

\BlankLine
\While{\colorbox{yellow!25}{$|L_R| < T_{rec}|R_E|$}}{
    \eIf{$|L_R| \geq 1$}{
        $L_{pre}\leftarrow Presume(L,\neg L)$\;
        $CL\leftarrow Train(L \cup L_{pre})$\;
        \tcp{Estimate \#relevant papers}
        \colorbox{yellow!25}{$R_E,Y\leftarrow SEMI(CL,E,L,L_R)$}\;
        $x\leftarrow Query(CL,\neg L,L_R)$\;
    }{
        $x\leftarrow argsort(BM25(\neg L,Q))[0]$\;
    }
    $L_R,L\leftarrow Include(x,R,L_R,L)$\;
    $\neg L\leftarrow E \setminus L$\;
}
\Return{$L_R$}\;
\BlankLine
\Fn{SEMI ($CL,E,L,L_R$)}{
    $|R_E|_{last}\leftarrow 0$\;
    $\neg L \leftarrow E \setminus L$\;
    \BlankLine
    \ForEach{$x \in E$}{
        $D(x) \leftarrow CL.decision\_function(x)$\;
        \If{$x \in |L_R|$}{
            $Y(x)\leftarrow 1 $\;
        }
        \Else{
            $Y(x)\leftarrow 0 $\;
        }
    }
    \BlankLine
    $|R_E| \leftarrow \sum\limits_{x\in E} Y(x)$\;
    \BlankLine
    \While{$|R_E|\neq |R_E|_{last}$}{
    \BlankLine
       \tcp{Fit and transform Logistic Regression}
       $LReg \leftarrow LogisticRegression(D,Y)$\;
       \BlankLine
       $Y \leftarrow TemporaryLabel(LReg,\neg L,Y)$\;
       \BlankLine
       $|R_E|_{last}\leftarrow |R_E|$\;
       \BlankLine
       \tcp{Estimation based on temporary labels}
       $|R_E| \leftarrow \sum\limits_{x\in E} Y(x)$\;
    }
    \Return{$|R_E|,Y$}\;
}
\BlankLine
\Fn{TemporaryLabel ($LReg,\neg L,Y$)}{
    $count \leftarrow 0$\;
    $target \leftarrow 1$\;
    $can \leftarrow \emptyset$\;
    \BlankLine
    \tcp{Sort $\neg L$ by descending order of $LReg(x)$}
    $\neg L \leftarrow SortBy(\neg L,LReg)$\;
    \BlankLine
    \ForEach{$x \in \neg L$}{
        $count \leftarrow count+LReg(x)$\;
        $can \leftarrow can \cup \{x\}$\;
        \If{$count \geq target$}{
            $Y(can[0]) \leftarrow 1$\;
            $target \leftarrow target+1$\;
            $can \leftarrow \emptyset$\;
        }
    }
    \Return{$Y$}\;
}

\caption{Psuedo Code for SEMI on FAST$^2$}\label{alg:alg3}
\end{algorithm}

To summarize this section:

\begin{RQ}{Answer to RQ2 ``When to stop?''}
    With our proposed estimator SEMI, we suggest the review to stop when $|L_R|\geq T_{rec}|R_E|$. We have shown in our results that the proposed stopping rule can stop the review close to the target recall.
\end{RQ}

The updated algorithm with {SEMI} as stopping rule is shown in Algorithm~\ref{alg:alg3}. Rows colored in
yellow
are the differences from Algorithm~\ref{alg:alg2}. Functions already described in Algorithm~\ref{alg:alg2} and \ref{alg:alg3} are omitted.

\section{How to correct}
\label{sect: How to correct}

This section drops the assumption of \emph{Reviewer never make mistakes} on top of Algorithm~\ref{alg:alg3} and explores this question:

{\bf RQ3: How to correct?} Human makes mistakes, and those mistakes can be amplified by misleading the machine learning model to train on the wrongly labeled papers. How to correct human error in time and accurately is a huge challenge. 

\subsection{Related Work}
\label{sect: How to correct: Related Works}

The observation of human errors in literature review is prevalent, e.g., \citet{wohlin2013reliability} documented that reviewers will not find same papers even with thorough reviews, and \citet{Cormack2017Navigating} concluded that a human
reviewer could achieve on the order of 70\% recall and 70\% precision. The human error becomes an even more severe problem when active learning is applied to assist the review since the active learning model would be misled by the mislabeled training examples~\citep{Voorhees2000Variations}. Therefore correcting these human errors becomes a crucial task for this work.

Via our literature review, we found two state of the art error correction methods that can be applied to solve this problem:
\begin{itemize}
\item
Kuhrmann'17: One simple way to correct human errors is by majority vote. As \citet{Kuhrmann2017On} described in 2017, simplest form of majority vote requires every paper to be reviewed by two different reviewers, and when the two reviewers do not label it consistently, a third reviewer is asked to make the final decision. The advantage of this method is that the human errors are corrected immediately and will not mislead the active learning model. However, this method requires at least three different reviewers and will at least double the review effort.
\item
Cormack'17: This error correction method~\citep{Cormack2017Navigating} is built upon the Cormack'16 knee stopping rule~\citep{Cormack2016Engineering}. After the review stops, papers labeled as ``relevant'' but reviewed after the inflection point ($x>i$) and papers labeled as ``non-relevant'' but reviewed before the inflection point ($x<i$) are sent to reviewers for a recheck. Comparing to {Kuhrmann'17}, this method requires much less extra review effort on rechecking, but can only correct human errors after the review, which leads to bias in the active learning model.
\end{itemize}


\subsection{Method}
\label{sect: How to correct: Method}

Taking into consideration the advantages and disadvantages of the existing error correction methods, we utilize the class probability estimation from SEMI to build a new error correction method called Disagree which now and then (whenever 50 new papers are labeled) rechecks some of the labeled papers. To improve the efficiency of the error correction, only those papers with labels that the current active learner disagrees most on are rechecked. Since human error rate on mislabeling relevant as non-relevant is much higher than vice versa (thus creating say 70\% precision and 70\% recall on highly imbalance dataset), rechecking effort is focused more on papers labeled as non-relevant. 

In this way, Disagree can 1) correct human errors in time; and 2) avoid wasting too much effort on rechecking correctly labeled papers. Algorithm~\ref{alg:alg4} shows how Disagree works in detail where the \emph{IncludeError} function simulates a human reviewer with a precision of $Prec$ and recall of $Rec$.

\begin{algorithm}[!htbp]
\scriptsize
\SetKwInOut{Input}{Input}
\SetKwInOut{Output}{Output}
\SetKwInOut{Parameter}{Parameter}
\SetKwRepeat{Do}{do}{while}
\Input{$E$, set of all candidate papers\\$R$, set of ground truth relevant papers\\$Q$, search query for BM25\\$T_{rec}$, target recall (default=0.95)}
\Output{$L_R$, set of included papers}
\BlankLine

$L\leftarrow \emptyset$\; $L_R\leftarrow \emptyset$\; $\neg L\leftarrow E$\; 
$|R_E|\leftarrow \infty$\;
\colorbox{yellow!25}{$Fixed\leftarrow \emptyset$}\;

\BlankLine
\While{$|L_R| < T_{rec}|R_E|$}{
    \eIf{$|L_R| \geq 1$}{
        $L_{pre}\leftarrow Presume(L,\neg L)$\;
        $CL\leftarrow Train(L \cup L_{pre})$\;
        $R_E,Y\leftarrow SEMI(CL,E,L,L_R)$\;
        \If{$|L| \bmod 50 == 0$}{
            \tcc{Check labeled papers which human and machine disagree most}
            \colorbox{yellow!25}{$L_R,L\leftarrow Disagree(CL,L,L_R)$}\;
        }
        $x\leftarrow Query(CL,\neg L,L_R)$\;
    }{
        $x\leftarrow argsort(BM25(\neg L,Q))[0]$\;
    }
    \tcp{Simulate review with human errors}
    \colorbox{yellow!25}{$L_R,L\leftarrow IncludeError(x,R,L_R,L)$}\;
    $\neg L\leftarrow E \setminus L$\;
}
\Return{$L_R$}\;
\BlankLine
\Fn{IncludeError($x,R,L_R,L$)}{
    \tcp{Simulate human errors with precision and recall}
    $Prec = 0.70$\;
    $Rec = 0.70$\;
    \eIf{$x \in L$}{
        $Fixed\leftarrow Fixed \cup x$\;
    }{
        $L\leftarrow L \cup x$\;
    }
    \eIf{$\big(x\in R$ \textbf{and} $random()< Rec\big)$ \textbf{or} $\big( x\notin R$ \textbf{and} $random()< \frac{|R|}{|E|-|R|}(\frac{Rec}{Prec}-Prec)\big)$}{
        $L_R\leftarrow L_R \cup x$\;
    }{
        $L_R\leftarrow L_R \setminus x$\;
    }
    \Return{$L_R$, $L$}\;
}
\BlankLine
\Fn{Disagree($CL,L,L_R,Fixed$)}{
    \tcp{Remove fixed items from checklist}
    $Check_I\leftarrow L\setminus L_R\setminus Fixed$\;
    $Check_R\leftarrow L_R \setminus Fixed$\;
    $Threshold \leftarrow 1.6/(1+CL.class\_weight)$\;
    \tcc{Rank papers by the level of human-machine disagreements}
    $R_{dis} \leftarrow argsort(CL.predict\_proba(Check_R)<Threshold$\;
    $I_{dis} \leftarrow argsort(CL.predict\_proba(Check_I)>Threshold$\;
    \tcp{Ask human to relabel}
    \ForEach{$x\in R_{dis}\cup I_{dis}$}{
        $L_R,L\leftarrow IncludeError(x,R,L_R,L)$\;
    }
    \Return{$L_R$, $L$}\;
}
\caption{Psuedo Code for Disagree on FAST$^2$}\label{alg:alg4}
\end{algorithm}

\subsection{Experiments}
\label{sect: How to correct: Experiments}

In this subsection, we design experiments to answer \textbf{RQ3} by comparing Disagree with:
\begin{itemize}
\item
None: i.e. no error correction (so just Algorithm~\ref{alg:alg3} with human errors); 
\item
Kuhrmann'17: majority vote with three reviewers of same error rate; and \item
Cormack'17: knee method with $\rho=6$. 
\end{itemize}
Our simulations are conducted on the four SE literature review datasets
of Table~\ref{tab: stats} with increasing human error rate injected (from 100\% recall and 100\% precision to 70\% recall and 70\% precision).

Because of the introduction of human errors, $L_R \neq L\cap R$ anymore. We use $tp=|R \cap L_R|$ to represent the true positive, and $ct$ to represent the actual review effort ($ct\leftarrow ct+1$ each time a new/labeled paper is reviewed/rechecked). Performance of each method is assessed by the following metrics:
\begin{itemize}
\item
{Recall}: $tp/|R|$ measures the percentage of relevant papers being included in the final inclusion list. The Higher, the better. 
\item
{Precision}: $tp/|L_R|$ measures that in the set of papers labeled as relevant by human reviewers, how many of them are relevant. The higher, the better. 
\item
{Effort}: $ct$ measures the actual number of papers being reviewed by human reviewers (including rechecking). The lower, the better.
\end{itemize}

\begin{table*}[!htb]
\caption{Comparison of different error correction methods.
}
\label{tab: errors}
\begin{center}
\begin{threeparttable}
\small
\setlength\tabcolsep{3pt}

\begin{tabular}{ll|rrr|rrrr|rrrr}
                             &                  & \multicolumn{2}{c}{Recall} & \textbf{b/a}                                      &  & \multicolumn{2}{c}{Precision} & \textbf{d/c}                                       &  & \multicolumn{2}{c}{Effort} & \textbf{f/e}                                      \\
                             &                  & a=100\%       & b=70\%      &                                   &  & c= 100\%       & d=70\%       &                                   &  & e=100\%      & f=70\%      &                            \\ \hline
\multirow{5}{*}{None}        & Wahono           & 95(0)          & 69(8)        & 73(8)                                                &  & 100(0)          & 90(4)         & 90(4)                                                 &  & 1,180(67)        & 1,640(335)       & 139(34)                                               \\
                             & Hall             & 98(0)          & 70(4)        & 71(4)                                                &  & 100(0)          & 95(3)         & 95(3)                                                 &  & 500(30)          & 1,480(465)       & 296(100)                                               \\
                             & Radjenovi{\'c} & 94(0)          & 66(6)        & 70(7)                                                &  & 100(0)          & 90(3)         & 90(3)                                                 &  & 760(37)          & 1,405(467)       & 185(66)                                               \\
                             & Kitchenham       & 91(2)          & 69(4)        & 76(7)                                                &  & 100(0)          & 86(6)         & 86(6)                                                 &  & 510(20)          & 660(140)         & 129(31)                                               \\
                             &                  &               &             & \cellcolor{lightgray}\textcolor{black}{73} &  &                &              & \cellcolor{lightgray}\textcolor{black}{90}  &  &              &             & \cellcolor{lightgray}\textcolor{black}{187} \\ \hline
\multirow{5}{*}{Kuhrmann'17} & Wahono           & 95(0)          & 75(7)        & 79(7)                                                &  & 100(0)          & 100(0)        & 100(0)                                                &  & 2,380(140)        & 3,193(471)       & 134(26)                                               \\
                             & Hall             & 98(0)          & 77(5)        & 79(5)                                                &  & 100(0)          & 100(0)        & 100(0)                                                &  & 980(75)          & 1,603(525)       & 164(61)                                               \\
                             & Radjenovi{\'c} & 94(0)          & 76(8)        & 81(9)                                                &  & 100(0)          & 100(0)        & 100(0)                                                &  & 1,520(80)        & 2,223(788)       & 146(57)                                               \\
                             & Kitchenham       & 91(0)          & 73(4)        & 80(5)                                               &  & 100(0)          & 100(0)        & 100(0)                                                &  & 1,020(20)        & 1,208(175)       & 118(20)                                               \\
                             &                  &               &             & \cellcolor{lightgray}\textcolor{black}{80} &  &                &              & \cellcolor{lightgray}\textcolor{black}{100} &  &              &             & \cellcolor{lightgray}\textcolor{black}{140} \\ \hline
\multirow{5}{*}{Cormack'17}  & Wahono           & 94(7)          & 65(12)        & 69(20)                                               &  & 100(0)          & 99(4)        & 99(4)                                                 &  & 999(238)          & 712(360)         & 71(60)                                                \\
                             & Hall             & 97(1)          & 84(4)        & 87(5)                                                &  & 100(0)          & 99(2)         & 99(2)                                                 &  & 552(58)          & 660(137)         & 120(35)                                              \\
                             & Radjenovi{\'c} & 86(4)          & 75(10)        & 87(16)                                                &  & 100(0)          & 98(3)         & 98(3)                                                 &  & 544(114)          & 713(275)         & 131(74)                                               \\
                             & Kitchenham       & 96(4)          & 87(14)        & 90(19)                                                &  & 100(0)          & 85(11)         & 85(11)                                                 &  & 1041(168)         & 1910(1004)        & 183(113)                                               \\
                             &                  &               &             & \cellcolor{lightgray}\textcolor{black}{83} &  &                &              & \cellcolor{lightgray}\textcolor{black}{95}  &  &              &             & \cellcolor{lightgray}\textcolor{black}{126} \\ \hline
\multirow{5}{*}{Disagree}    & Wahono           & 95(0)          & 88(3)        & 93(3)                                                &  & 100(0)          & 90(5)         & 90(5)                                                 &  & 2,051(63)        & 2,531(406)       & 123(23)                                               \\
                             & Hall             & 98(0)          & 90(3)        & 92(3)                                                &  & 100(0)          & 98(3)         & 98(3)                                                 &  & 755(51)          & 897(206)         & 119(34)                                               \\
                             & Radjenovi{\'c} & 94(0)          & 88(6)        & 94(6)                                                &  & 100(0)          & 91(3)         & 91(3)                                                 &  & 1,390(89)        & 1,577(336)       & 113(31)                                               \\
                             & Kitchenham       & 91(2)          & 81(7)        & 89(10)                                                &  & 100(0)          & 85(7)         & 85(7)                                                 &  & 940(30)          & 975(144)         & 104(19)                                               \\
                             &                  &               &             & \cellcolor{lightgray}\textcolor{black}{92} &  &                &              & \cellcolor{lightgray}\textcolor{black}{91}  &  &              &             & \cellcolor{lightgray}\textcolor{black}{115} \\ \hline
\end{tabular}

\begin{tablenotes}\small
Columns a, c, e report results from the {\em optimistic case} i.e., when relevancy is assessed 100\% correctly. Columns b, d, f report results from the more
{\em realistic case} where human oracles have 70\% recall and 70\% precision.
Each experiment is repeated for 30 times. Results are shown in this table as median(IQR) in percentage. Cormack'17 uses knee stopping rule with $\rho=6$ while others use SEMI stopping rule with target recall $T_{rec}=0.95$.  The column \textbf{recall} stands for $tp/|R|$, \textbf{precision} stands for $tp/|L_R|$, and \textbf{effort} $ct$ is the number of papers being reviewed (rechecking a paper also increase the \textbf{effort} by 1).
Cells in gray show the average of that treatment across the four datasets, for recall and precision, higher is better, while for effort, lower is better.
\end{tablenotes}
\end{threeparttable}\end{center}
\end{table*}

\subsection{Results}
\label{sect: How to correct: Results}

Table~\ref{tab: errors} shows the median results for 30 repeated simulations.  While we collected results for human reviewer recall, precision $\in \{100\%,90\%,80\%,70\%\}$, the trend across all the results was apparent. Hence, for the sake of brevity, 
we just discuss the 100\% and 70\% results. Full results are available at https://tiny.cc/error\_fast2. Note that:
\begin{itemize}
\item
For precision and recall, better tactics have less  deterioration as errors increase; i.e. {\em higher} b/a and d/c values are {\em better}.
\item
For effort (which measures the number of papers reviewed/rechecked), better tactics demand fewer papers are read; i.e. {\em lower} f/e values are {\em better}.
\item
The cells in gray show median changes over the four data sets for all the 
error correction tactics.
\end{itemize}

Of the results in Table~\ref{tab: errors}, the None tactic (which does not feature any error correction) is the worst. While it paints good precision, as errors are added to the labeling process, recall decreases to 69\% and  53\% more efforts need to be spent on reviewing.

Kuhrmann'17 is the go-to treatment if 100\% precision is required. The downside is that it usually requires most effort to achieve the 100\% precision and its recall is second worst among all four treatments. In general, we do not suggest Kuhrmann'17 since recall is most important.

Cormack'17 is the least stable treatment among the four. On Wahono, Hall, Radjenovi{\'c} datasets, it uses the least effort but achieves very poor recall; while on Kitchenham dataset, it achieves highest recall but also costs most effort. 

Turning now to the {\em  Disagree} results, we see that this tactic maintains high recall (92\% of their original values in average) with acceptable precision (91\% of their original values in average) at the cost of modest increases in effort (115\% of their original values in average). Comparing to the {\em optimistic case}, Disagree retains high recall when human error increases with the cost of around 100\% extra effort.

\begin{RQ}{Answer to RQ3 ``How to Correct?"}
   If reviewers frequently pause and recheck old conclusions whose labels contradict the current conclusion of the learner, the performance deterioration associated with human error can be alleviated without much overhead rechecking effort.
\end{RQ}

The updated algorithm with {Disagree} to correct human misclassifications is shown in Algorithm~\ref{alg:alg4}. Rows colored in
yellow are the differences from Algorithm~\ref{alg:alg3}. Functions already described in Algorithm~\ref{alg:alg3} are omitted.

\section{Discussion}
\label{sect: discussion}

\subsection{FAST$^2$}
\label{sect: fast2}
\begin{figure}[!t]
    \centering
    \includegraphics[width=\linewidth]{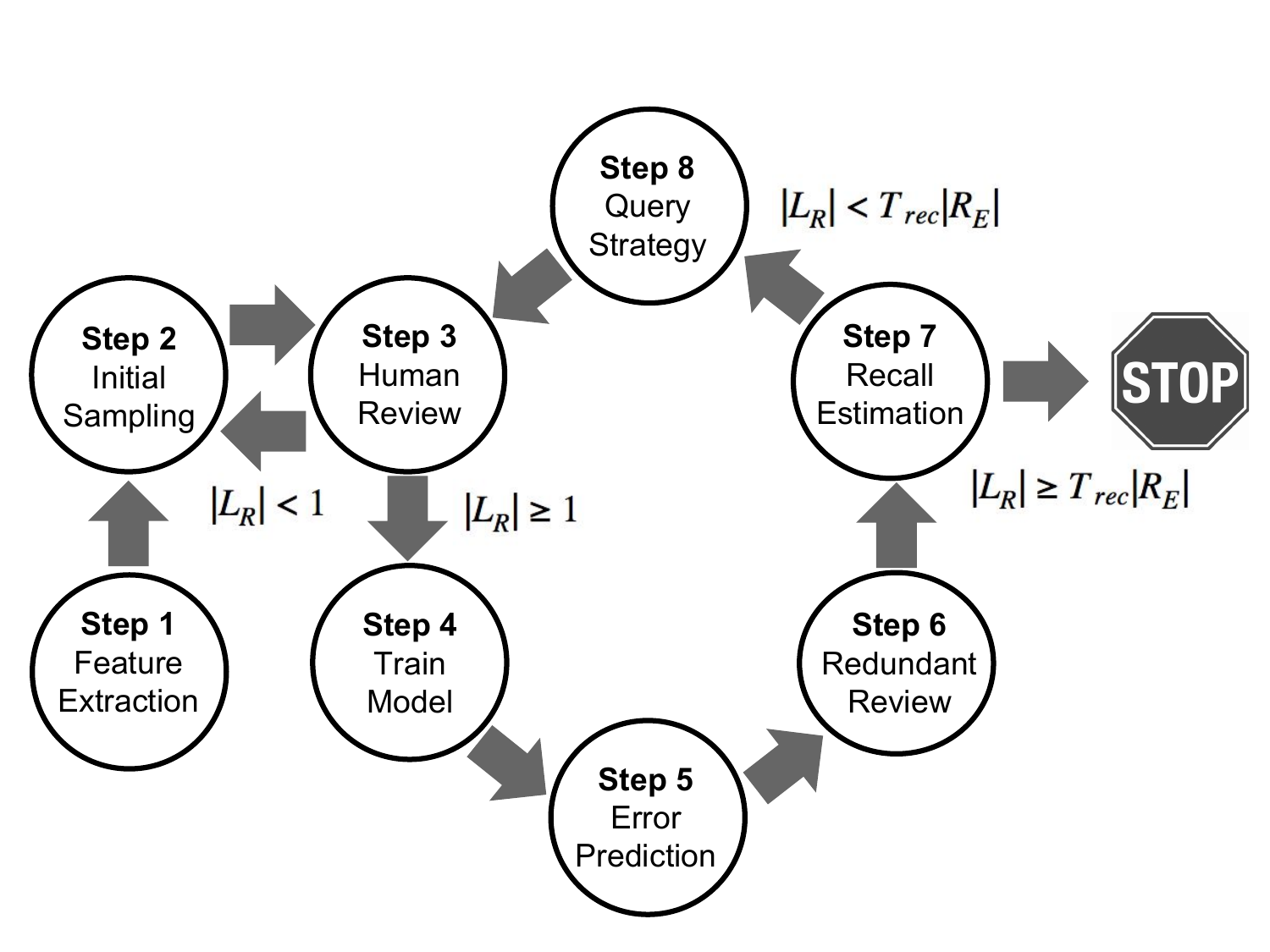}
    \caption{The FAST$^2$ framework.}
    \label{fig: framework}
\end{figure}

Combining the results above from Section~\ref{sect: How to Start} to Section~\ref{sect: How to correct}, we propose our new reading tool FAST$^2$ which is built on top of FASTREAD~\citep{yu2018finding} and its framework is shown in Figure~\ref{fig: framework} with the following features:
\begin{itemize}
\item
\textbf{Initial Sampling} with BM25 ranking keyword search.
\item
\textbf{Recall Estimation} by SEMI estimator.
\item
\textbf{Error Prediction} by Disagree.
\end{itemize}
When applied to other domains, same framework should be followed while detailed techniques can be changed or adapted. The new FAST$^2$ tool is available on Zenodo~\citep{fastread} and is continuously updated on Github~\footnote{https://github.com/fastread/src}.

\subsection{Threats to Validity}
 \label{sect: Threats to Validity}

There are several validity threats to the design of this study~\citep{feldt2010validity}. Any conclusions made from this work must be considered with the following issues in mind:

\textbf{Conclusion validity} focuses on the significance of the treatment. It applies to our analysis of which treatment works best for ``how to start''. Although the suggested treatment (Rank-BM25, SEMI, Disagree) performs consistently better than {FASTREAD}, it is not always the best treatment on every dataset. 

\textbf{Internal validity }focuses on how sure we can be that the treatment caused the outcome. This applies to our analysis of which treatment works best for ``how to start'' since the domain knowledge applied is decided by the authors and different domain knowledge might lead to different outcomes.

\textbf{Construct validity }focuses on the relation between the theory behind the experiment and the observation. In this work, we evaluate different treatments with a target of reaching 0.95 recall. Although it is right now the widely accepted target for active learning based literature reviews~\citep{cohen2011performance,o2015using}, either increasing or decreasing the required final recall may result in a different ranking of treatments.

\textbf{External validity} concerns how well the conclusion can be applied outside. All the conclusions in this study are drawn from the experiments running on four software engineering systematic literature review datasets generated from \citet{hall2012systematic,wahono2015systematic,radjenovic2013software,kitchenham2013systematic}. Therefore, such conclusions may not be applicable to datasets of different scenarios, e.g., citation screening from evidence-based medicine or TAR from e-discovery. Such bias threatens any classification experiment. The best any researcher can do is to document that bias then makes available to the general research community all the materials used in a study (with the hope that other researchers will explore similar work on different datasets). To this end, we have published all our code and data on Zenodo~\citep{fastread}. Also, there are parameters decided as engineering judgment which may not be the best choice. Tuning/tweaking such parameters might lead to different conclusions. Finally, this concern also applies to the results of ``when to stop'' and ``How to correct'' since the task of selecting relevant papers is multi-objective, and we choose SEMI and Disagree as for the best methods in favor of their trade-off between recall and review effort. In other circumstances, a user may prefer other methods in favor of their own goals.

\section{Conclusions}
\label{sect: Conclusion}

Unless other people can find our research, we will risk our work being unacknowledged, uncited, and forgotten. One significant barrier to finding related research is the effort of selecting papers to read. A systematically scheduled review usually requires researchers to review thousands of papers to find the dozens of papers relevant to their research and this usually costs weeks to months of work. Previously, we have built a state of the art reading support tools to reduce that effort by applying active learning methods~\citep{yu2018finding}. 

Although our reading tool FASTREAD achieved a good reduction on review efforts, it was not ready for practical use because of its three unrealistic assumptions: 1) no external domain knowledge, 2) reviewer knows when to stop reviewing, and 3) reviewer never make mistakes. Accordingly, in this paper, we extend our previous work by dropping and addressing the above three assumptions with different techniques either found from existing works or created by ourselves. 

Our results suggested that 1) with a little domain knowledge (two or three keywords or data from previous reviews), we can further reduce the review effort by 10\% to 20\% with much better robustness. 2) By training a logistic regression model with semi-supervised learning, we can estimate the number of relevant papers accurately (better than the state of the art estimator Wallace'13~\citep{wallace2013active}) in an early stage, thus providing a reliable stopping rule (which is better than state of the art stopping rules in literature). 3) By asking the reviewer to recheck those papers whose labels the active learner disagrees most on, human errors can be successfully corrected without much extra review cost (proved to be a better error correction method than the state of the art ones in literature). 

Based on these results, we build our new tool FAST$^2$ as represented in Algorithm~\ref{alg:alg4}. It uses keywords search and BM25 ranking for finding the first relevant example, trains the {SEMI} estimator, rechecks labeled papers now and then, and stops review when $|L_R|\geq T_{rec}|R_E|$. 

Considering the problems raised in Section~\ref{sect: Threats to Validity}, our future works will focus on the following aspects:
\begin{itemize}
\item
{\em Our conclusions are drawn from only four SE literature review datasets, which may incur sampling bias.} In future work, we will validate the results on more datasets including those from medical and legal domains.
\item
{\em Currently the target for active learning based review is to achieve 95\% recall.} It is an open and interesting problem whether there exists an efficient way to retrieve the rest 5\% relevant papers.
\item
{\em The magic parameters are selected based on expert suggestions.} Tuning is challenging for this active learning schema since the available labeled data are limited and can only be obtained at cost. How to design a feasible tuning schema with the limited resource and how to avoid overfitting can be difficult problems but solving such problems provides great value to the domain. 

\end{itemize}


\section*{Acknowledgement}
The authors thank Barbara Kitchenham for
her attention to this work and for
sharing with us the ``Kitchenham'' dataset used in these experiments.

\vspace*{0.5mm}

\balance

\bibliographystyle{elsarticle-num}

\begin{thebibliography}{}

\bibitem[Adeva et~al., 2014]{adeva2014automatic}
Adeva, J.~G., Atxa, J.~P., Carrillo, M.~U., and Zengotitabengoa, E.~A. (2014).
\newblock Automatic text classification to support systematic reviews in
  medicine.
\newblock {\em Expert Systems with Applications}, 41(4):1498--1508.

\bibitem[Bowes et~al., 2012]{bowes2012slurp}
Bowes, D., Hall, T., and Beecham, S. (2012).
\newblock Slurp: a tool to help large complex systematic literature reviews
  deliver valid and rigorous results.
\newblock In {\em Proceedings of the 2nd international workshop on Evidential
  assessment of software technologies}, pages 33--36. ACM.

\bibitem[Carver et~al., 2013]{carver2013identifying}
Carver, J.~C., Hassler, E., Hernandes, E., and Kraft, N.~A. (2013).
\newblock Identifying barriers to the systematic literature review process.
\newblock In {\em 2013 ACM/IEEE International Symposium on Empirical Software
  Engineering and Measurement}, pages 203--212. IEEE.

\bibitem[Chapelle et~al., 2017]{Chapelle2017Introduction}
Chapelle, O., Schã¶Lkopf, B., and Zien, A. (2017).
\newblock Introduction to semi-supervised learning.
\newblock {\em Journal of the Royal Statistical Society}, 172(2):1826--1831.

\bibitem[Cliff, 1993]{cliff1993dominance}
Cliff, N. (1993).
\newblock Dominance statistics: Ordinal analyses to answer ordinal questions.
\newblock {\em Psychological Bulletin}, 114(3):494.

\bibitem[Cohen, 2011]{cohen2011performance}
Cohen, A.~M. (2011).
\newblock Performance of support-vector-machine-based classification on 15
  systematic review topics evaluated with the wss@ 95 measure.
\newblock {\em Journal of the American Medical Informatics Association},
  18(1):104--104.

\bibitem[Cohen et~al., 2006]{cohen2006reducing}
Cohen, A.~M., Hersh, W.~R., Peterson, K., and Yen, P.-Y. (2006).
\newblock Reducing workload in systematic review preparation using automated
  citation classification.
\newblock {\em Journal of the American Medical Informatics Association},
  13(2):206--219.

\bibitem[Cormack and Grossman, 2014]{cormack2014evaluation}
Cormack, G.~V. and Grossman, M.~R. (2014).
\newblock Evaluation of machine-learning protocols for technology-assisted
  review in electronic discovery.
\newblock In {\em Proceedings of the 37th international ACM SIGIR conference on
  Research \& development in information retrieval}, pages 153--162. ACM.

\bibitem[Cormack and Grossman, 2015]{cormack2015autonomy}
Cormack, G.~V. and Grossman, M.~R. (2015).
\newblock Autonomy and reliability of continuous active learning for
  technology-assisted review.
\newblock {\em arXiv preprint arXiv:1504.06868}.

\bibitem[Cormack and Grossman, 2016a]{Cormack2016Engineering}
Cormack, G.~V. and Grossman, M.~R. (2016a).
\newblock Engineering quality and reliability in technology-assisted review.
\newblock In {\em Proceedings of the 39th International ACM SIGIR conference on
  Research and Development in Information Retrieval}, pages 75--84. ACM.

\bibitem[Cormack and Grossman, 2016b]{cormack2016scalability}
Cormack, G.~V. and Grossman, M.~R. (2016b).
\newblock Scalability of continuous active learning for reliable high-recall
  text classification.
\newblock In {\em Proceedings of the 25th ACM International on Conference on
  Information and Knowledge Management}, pages 1039--1048. ACM.

\bibitem[Cormack and Grossman, 2017]{Cormack2017Navigating}
Cormack, G.~V. and Grossman, M.~R. (2017).
\newblock Navigating imprecision in relevance assessments on the road to total
  recall: Roger and me.
\newblock In {\em The International ACM SIGIR Conference}, pages 5--14.

\bibitem[Di~Nunzio, 2018]{di2018study}
Di~Nunzio, G.~M. (2018).
\newblock A study of an automatic stopping strategy for technologically
  assisted medical reviews.
\newblock In {\em European Conference on Information Retrieval}, pages
  672--677. Springer.

\bibitem[Efron, 1982]{efron1982jackknife}
Efron, B. (1982).
\newblock {\em The jackknife, the bootstrap and other resampling plans}.
\newblock SIAM.

\bibitem[Feldt and Magazinius, 2010]{feldt2010validity}
Feldt, R. and Magazinius, A. (2010).
\newblock Validity threats in empirical software engineering research-an
  initial survey.
\newblock In {\em SEKE}, pages 374--379.

\bibitem[Felizardo et~al., 2016]{felizardo2016using}
Felizardo, K.~R., Mendes, E., Kalinowski, M., Souza, {\'E}.~F., and Vijaykumar,
  N.~L. (2016).
\newblock Using forward snowballing to update systematic reviews in software
  engineering.
\newblock In {\em Proceedings of the 10th ACM/IEEE International Symposium on
  Empirical Software Engineering and Measurement}, page~53. ACM.

\bibitem[Felizardo et~al., 2010]{felizardo2010approach}
Felizardo, K.~R., Nakagawa, E.~Y., Feitosa, D., Minghim, R., and Maldonado,
  J.~C. (2010).
\newblock An approach based on visual text mining to support categorization and
  classification in the systematic mapping.
\newblock In {\em Proc. of EASE}, volume~10, pages 1--10.

\bibitem[Goeuriot et~al., 2017]{clef2017}
Goeuriot, L., Kelly, L., Suominen, H., N{\'{e}}v{\'{e}}ol, A., Robert, A.,
  Kanoulas, E., Spijker, R., Palotti, J. R.~M., and Zuccon, G. (2017).
\newblock Clef 2017 ehealth evaluation lab overview.
\newblock In {\em Experimental {IR} Meets Multilinguality, Multimodality, and
  Interaction - 8th International Conference of the {CLEF} Association, {CLEF}
  2017, Dublin, Ireland, September 11-14, 2017, Proceedings}, Lecture Notes in
  Computer Science. Springer.

\bibitem[Grossman and Cormack, 2013]{grossman2013}
Grossman, M.~R. and Cormack, G.~V. (2013).
\newblock The grossman-cormack glossary of technology-assisted review with
  foreword by john m. facciola, u.s. magistrate judge.
\newblock {\em Federal Courts Law Review}, 7(1):1--34.

\bibitem[Hall et~al., 2012]{hall2012systematic}
Hall, T., Beecham, S., Bowes, D., Gray, D., and Counsell, S. (2012).
\newblock A systematic literature review on fault prediction performance in
  software engineering.
\newblock {\em IEEE Transactions on Software Engineering}, 38(6):1276--1304.

\bibitem[Hassler et~al., 2016]{hassler2016identification}
Hassler, E., Carver, J.~C., Hale, D., and Al-Zubidy, A. (2016).
\newblock Identification of {SLR} tool needs --- results of a community
  workshop.
\newblock {\em Information and Software Technology}, 70:122--129.

\bibitem[Hassler et~al., 2014]{hassler2014outcomes}
Hassler, E., Carver, J.~C., Kraft, N.~A., and Hale, D. (2014).
\newblock Outcomes of a community workshop to identify and rank barriers to the
  systematic literature review process.
\newblock In {\em Proceedings of the 18th International Conference on
  Evaluation and Assessment in Software Engineering}, page~31. ACM.

\bibitem[Jalali and Wohlin, 2012]{jalali2012systematic}
Jalali, S. and Wohlin, C. (2012).
\newblock Systematic literature studies: database searches vs. backward
  snowballing.
\newblock In {\em Proceedings of the ACM-IEEE international symposium on
  Empirical software engineering and measurement}, pages 29--38. ACM.

\bibitem[Kanoulas et~al., 2017]{task2}
Kanoulas, E., Li, D., Azzopardi, L., and Spijker, R. (2017).
\newblock Overview of the {CLEF} technologically assisted reviews in empirical
  medicine.
\newblock In {\em Working Notes of {CLEF} 2017 - Conference and Labs of the
  Evaluation forum, Dublin, Ireland, September 11-14, 2017.}, {CEUR} Workshop
  Proceedings. CEUR-WS.org.

\bibitem[Keele, 2007]{keele2007guidelines}
Keele, S. (2007).
\newblock Guidelines for performing systematic literature reviews in software
  engineering.
\newblock In {\em Technical report, Ver. 2.3 EBSE Technical Report. EBSE}.

\bibitem[Kitchenham, 2004]{kitchenham2004procedures}
Kitchenham, B. (2004).
\newblock Procedures for performing systematic reviews.
\newblock {\em Keele, UK, Keele University}, 33(2004):1--26.

\bibitem[Kitchenham and Brereton, 2013]{kitchenham2013systematic}
Kitchenham, B. and Brereton, P. (2013).
\newblock A systematic review of systematic review process research in software
  engineering.
\newblock {\em Information and software technology}, 55(12):2049--2075.

\bibitem[Kitchenham et~al., 2004]{kitchenham2004evidence}
Kitchenham, B.~A., Dyba, T., and Jorgensen, M. (2004).
\newblock Evidence-based software engineering.
\newblock In {\em Proceedings of the 26th international conference on software
  engineering}, pages 273--281. IEEE Computer Society.

\bibitem[Kuhrmann et~al., 2017]{Kuhrmann2017On}
Kuhrmann, M., Fernández, D.~M., and Daneva, M. (2017).
\newblock On the pragmatic design of literature studies in software
  engineering: an experience-based guideline.
\newblock {\em Empirical Software Engineering}, 22(6):2852--2891.

\bibitem[Liu et~al., 2016]{liu2016comparative}
Liu, J., Timsina, P., and El-Gayar, O. (2016).
\newblock A comparative analysis of semi-supervised learning: The case of
  article selection for medical systematic reviews.
\newblock {\em Information Systems Frontiers}, pages 1--13.

\bibitem[Malheiros et~al., 2007]{malheiros2007visual}
Malheiros, V., Hohn, E., Pinho, R., Mendonca, M., and Maldonado, J.~C. (2007).
\newblock A visual text mining approach for systematic reviews.
\newblock In {\em First International Symposium on Empirical Software
  Engineering and Measurement (ESEM 2007)}, pages 245--254. IEEE.

\bibitem[Marshall et~al., 2015]{marshall2015tools}
Marshall, C., Brereton, P., and Kitchenham, B. (2015).
\newblock Tools to support systematic reviews in software engineering: A
  cross-domain survey using semi-structured interviews.
\newblock In {\em Proceedings of the 19th International Conference on
  Evaluation and Assessment in Software Engineering}, page~26. ACM.

\bibitem[Mittas and Angelis, 2013]{mittas2013ranking}
Mittas, N. and Angelis, L. (2013).
\newblock Ranking and clustering software cost estimation models through a
  multiple comparisons algorithm.
\newblock {\em IEEE Transactions on software engineering}, 39(4):537--551.

\bibitem[Miwa et~al., 2014]{miwa2014reducing}
Miwa, M., Thomas, J., O'Mara-Eves, A., and Ananiadou, S. (2014).
\newblock Reducing systematic review workload through certainty-based
  screening.
\newblock {\em Journal of biomedical informatics}, 51:242--253.

\bibitem[O'Mara-Eves et~al., 2015]{o2015using}
O'Mara-Eves, A., Thomas, J., McNaught, J., Miwa, M., and Ananiadou, S. (2015).
\newblock Using text mining for study identification in systematic reviews: a
  systematic review of current approaches.
\newblock {\em Systematic reviews}, 4(1):5.

\bibitem[Ouzzani et~al., 2016]{Ouzzani2016}
Ouzzani, M., Hammady, H., Fedorowicz, Z., and Elmagarmid, A. (2016).
\newblock Rayyan---a web and mobile app for systematic reviews.
\newblock {\em Systematic Reviews}, 5(1):210.

\bibitem[Paynter et~al., 2016]{paynter2016epc}
Paynter, R., Ba{\~n}ez, L.~L., Berliner, E., Erinoff, E., Lege-Matsuura, J.,
  Potter, S., and Uhl, S. (2016).
\newblock Epc methods: an exploration of the use of text-mining software in
  systematic reviews.
\newblock {\em Research White Paper}.

\bibitem[Radjenovi{\'c} et~al., 2013]{radjenovic2013software}
Radjenovi{\'c}, D., Heri{\v{c}}ko, M., Torkar, R., and {\v{Z}}ivkovi{\v{c}}, A.
  (2013).
\newblock Software fault prediction metrics: A systematic literature review.
\newblock {\em Information and Software Technology}, 55(8):1397--1418.

\bibitem[Robertson et~al., 2009]{robertson2009probabilistic}
Robertson, S., Zaragoza, H., et~al. (2009).
\newblock The probabilistic relevance framework: Bm25 and beyond.
\newblock {\em Foundations and Trends{\textregistered} in Information
  Retrieval}, 3(4):333--389.

\bibitem[Roegiest et~al., 2015]{roegiest2015trec}
Roegiest, A., Cormack, G.~V., Grossman, M., and Clarke, C. (2015).
\newblock Trec 2015 total recall track overview.
\newblock {\em Proc. TREC-2015}.

\bibitem[Ros et~al., 2017]{ros2017machine}
Ros, R., Bjarnason, E., and Runeson, P. (2017).
\newblock A machine learning approach for semi-automated search and selection
  in literature studies.
\newblock In {\em Proceedings of the 21st International Conference on
  Evaluation and Assessment in Software Engineering}, pages 118--127. ACM.

\bibitem[Scott and Knott, 1974]{scott1974cluster}
Scott, A.~J. and Knott, M. (1974).
\newblock A cluster analysis method for grouping means in the analysis of
  variance.
\newblock {\em Biometrics}, pages 507--512.

\bibitem[Shemilt et~al., 2016]{shemilt2016use}
Shemilt, I., Khan, N., Park, S., and Thomas, J. (2016).
\newblock Use of cost-effectiveness analysis to compare the efficiency of study
  identification methods in systematic reviews.
\newblock {\em Systematic reviews}, 5(1):140.

\bibitem[Thomas et~al., 2010]{thomas2010eppi}
Thomas, J., Brunton, J., and Graziosi, S. (2010).
\newblock Eppi-reviewer 4.0: software for research synthesis.

\bibitem[Umemoto et~al., 2016]{Umemoto2016ScentBar}
Umemoto, K., Yamamoto, T., and Tanaka, K. (2016).
\newblock Scentbar: A query suggestion interface visualizing the amount of
  missed relevant information for intrinsically diverse search.
\newblock In {\em International ACM SIGIR Conference on Research and
  Development in Information Retrieval}, pages 405--414.

\bibitem[Voorhees, 2000]{Voorhees2000Variations}
Voorhees, E.~M. (2000).
\newblock Variations in relevance judgments and the measurement of retrieval
  effectiveness.
\newblock {\em Information Processing \& Management}, 36(5):697--716.

\bibitem[Wahono, 2015]{wahono2015systematic}
Wahono, R.~S. (2015).
\newblock A systematic literature review of software defect prediction:
  research trends, datasets, methods and frameworks.
\newblock {\em Journal of Software Engineering}, 1(1):1--16.

\bibitem[Wallace and Dahabreh, 2012]{wallace2012class}
Wallace, B.~C. and Dahabreh, I.~J. (2012).
\newblock Class probability estimates are unreliable for imbalanced data (and
  how to fix them).
\newblock In {\em Data Mining (ICDM), 2012 IEEE 12th International Conference
  on}, pages 695--704. IEEE.

\bibitem[Wallace et~al., 2013a]{wallace2013active}
Wallace, B.~C., Dahabreh, I.~J., Moran, K.~H., Brodley, C.~E., and Trikalinos,
  T.~A. (2013a).
\newblock Active literature discovery for scoping evidence reviews: How many
  needles are there.
\newblock In {\em KDD workshop on data mining for healthcare (KDD-DMH)}.

\bibitem[Wallace et~al., 2013b]{wallace2013modernizing}
Wallace, B.~C., Dahabreh, I.~J., Schmid, C.~H., Lau, J., and Trikalinos, T.~A.
  (2013b).
\newblock Modernizing the systematic review process to inform comparative
  effectiveness: tools and methods.
\newblock {\em Journal of comparative effectiveness research}, 2(3):273--282.

\bibitem[Wallace et~al., 2012]{wallace2012deploying}
Wallace, B.~C., Small, K., Brodley, C.~E., Lau, J., and Trikalinos, T.~A.
  (2012).
\newblock Deploying an interactive machine learning system in an evidence-based
  practice center: abstrackr.
\newblock In {\em Proceedings of the 2nd ACM SIGHIT International Health
  Informatics Symposium}, pages 819--824. ACM.

\bibitem[Wallace et~al., 2010a]{wallace2010active}
Wallace, B.~C., Small, K., Brodley, C.~E., and Trikalinos, T.~A. (2010a).
\newblock Active learning for biomedical citation screening.
\newblock In {\em Proceedings of the 16th ACM SIGKDD international conference
  on Knowledge discovery and data mining}, pages 173--182. ACM.

\bibitem[Wallace et~al., 2011]{wallace2011should}
Wallace, B.~C., Small, K., Brodley, C.~E., and Trikalinos, T.~A. (2011).
\newblock Who should label what? instance allocation in multiple expert active
  learning.
\newblock In {\em SDM}, pages 176--187. SIAM.

\bibitem[Wallace et~al., 2010b]{wallace2010semi}
Wallace, B.~C., Trikalinos, T.~A., Lau, J., Brodley, C., and Schmid, C.~H.
  (2010b).
\newblock Semi-automated screening of biomedical citations for systematic
  reviews.
\newblock {\em BMC bioinformatics}, 11(1):1.

\bibitem[Wohlin, 2014]{wohlin2014guidelines}
Wohlin, C. (2014).
\newblock Guidelines for snowballing in systematic literature studies and a
  replication in software engineering.
\newblock In {\em Proceedings of the 18th international conference on
  evaluation and assessment in software engineering}, page~38. ACM.

\bibitem[Wohlin et~al., 2013]{wohlin2013reliability}
Wohlin, C., Runeson, P., Neto, P. A. d. M.~S., Engstr{\"o}m, E.,
  do~Carmo~Machado, I., and De~Almeida, E.~S. (2013).
\newblock On the reliability of mapping studies in software engineering.
\newblock {\em Journal of Systems and Software}, 86(10):2594--2610.

\bibitem[Yu et~al., 2018]{yu2018finding}
Yu, Z., Kraft, N.~A., and Menzies, T. (2018).
\newblock Finding better active learners for faster literature reviews.
\newblock {\em Empirical Software Engineering}, pages 1--26.

\bibitem[Yu and Menzies, 2017]{fastread}
Yu, Z. and Menzies, T. (2017).
\newblock fastread/src: Fast2.
\newblock https://doi.org/10.5281/zenodo.1184123.

\bibitem[Zhang et~al., 2011]{zhang2011empirical}
Zhang, H., Babar, M.~A., Bai, X., Li, J., and Huang, L. (2011).
\newblock An empirical assessment of a systematic search process for systematic
  reviews.
\newblock In {\em Evaluation \& Assessment in Software Engineering (EASE 2011),
  15th Annual Conference on}, pages 56--65. IET.

\end{thebibliography}

\end{document}